\documentclass[prb,twocolumn,amsfonts]{revtex4-1}
\usepackage{graphicx}
\usepackage{tikz,xcolor}
\usepackage{braket}
\usepackage{bm}
\usepackage{mathtools}
\usepackage[caption=false]{subfig}

\definecolor{mypink}{RGB}{255,0,127}
\definecolor{myblue}{RGB}{0,0,255}
\definecolor{mygreen}{RGB}{102,204,0}
\definecolor{myorange}{RGB}{255,128,0}
\definecolor{mypurple}{RGB}{127,0,255}

\begin{document}
\title{Conductance of gated junctions as a probe of topological interface states}

\date{\today}

\author{Eklavya Thareja}
\email{ethare1@lsu.edu}
\affiliation{Department of Physics and Astronomy, Louisiana State University, Baton Rouge, LA 70803}

\author{Ilya Vekhter}
\email{vekhter@lsu.edu}
\affiliation{Department of Physics and Astronomy, Louisiana State University, Baton Rouge, LA 70803}

\author{Mahmoud M. Asmar}
\email[Current address: Department of Physics and Astronomy, Center for Materials for Information Technology,
The University of Alabama, Tuscaloosa, AL 35487, USA.\ ]{masmar@ua.edu}
\affiliation{Department of Physics and Astronomy, Louisiana State University, Baton Rouge, LA 70803}

\begin{abstract}
Energy dispersion and spin orientation of the protected states at an interface between topological insulators (TIs) and non-topological materials depend on the charge redistribution, strain, and atomic displacement at the interface. Knowledge of these properties is essential for applications of topological compounds, but direct access to them in the interface geometry is difficult. We show that conductance of a gated double junction at the surface of a topological insulator exhibits oscillations and a quasi-linear decay as a function of gate voltage in different regimes. These give the values for the quasiparticle velocities along and normal to the junction in the interface region, and determine the symmetry of the topological interface states. The results are insensitive to the boundary conditions at the junction.

\end{abstract}

\maketitle

\section {Introduction} Topological materials~\cite{Rev_zhang, Rev_kane,Rev_hasan} acting as components of superconducting or spin-based electronic devices promise metrics exceeding current charge-based technologies~\cite{Mellnik2014,Wang:2017,Rakheja:2019}. These functionalities are enabled by the spin-momentum locking of the states localized at the interfaces between topologically trivial and non-trivial compounds. The details of this spin texture as a function of the in-plane momentum control the types of proximity-induced superconducting order~\cite{Fu_Kane_2008,Alicea2012}, spin accumulation under transport current~\cite{Burkov2010,Pesin2012}, and the gap opening under exchange fields~\cite{Rev_zhang,Alicea2012}. Proposed applications of topological insulators (TIs) commonly assume that the interface states have the same linear Dirac-like dispersion and helical spin texture 
as the states at vacuum TI termination~\cite{Rev_hasan}. Under this assumption, once material advances eliminated contribution of the bulk states and allowed tuning of the chemical potential~\cite{Zhang2011,Arakane2012}, the interface-based devices should have become  achievable.

However, topological interface states (TIS) are not guaranteed to be similar to the surface states. Generically, interfaces have lower symmetry than surfaces due to strain, lattice reconstruction, polar charge redistribution, broken bonds, \textit{etc}. Symmetry breaking (in the absence of fine tuning) simultaneously distorts the Dirac cone and breaks the helicity of the topological states inducing an out-of-plane spin component~\cite{FZhang2012,mahmoud_se_ti} thereby dramatically changing the response to the exchange field~\cite{mahmoud_se_ti} and types of the induced superconducting orders~\cite{Alspaugh2018}. Therefore, it is highly desirable to have straightforward tests of the symmetry breaking in the TIS.

Direct access (ARPES or scanning tunneling spectroscopy) to the TIS is very challenging due to the capping material, and indirect optical measurements (\textit{e.g.} Kerr rotation~\cite{Wang2017,Mondal2018}) are often hard to interpret.
Here we demonstrate that the dc conductance in a gated double junction setup for a topological heterostructure, shown in Fig.~\ref{fig:device}(a), exhibits a quasi-linear variation in one range, and Fabry-Perot-like oscillations in another range of gate voltages. The "kink" voltage at the edge of the linear regime and the period of the oscillations give the quasiparticle velocities parallel and normal to the junction, and hence the dispersion anisotropy  of the hidden topological interface state. Symmetry considerations supported by model calculations~\cite{FZhang2012,mahmoud_se_ti} dictate that this anisotropy is accompanied by the tilting of the spins out of the plane of the interface.  

{The experiment we propose does not quantify the degree to which spin is tilted away from the plane of the interface, and only detects if this tilting is present. However, its advantage is that it uses straightforward conductance measurements, and is robust with respect to the scattering at the junction, see below. Therefore we believe that the proposed method can be easily used to check whether the spins are helical and whether the junction should exhibit behavior suggested by simplified models. In addition, our analysis of the role of junction scattering offers a clear perspective on the role of this, inevitably present, phenomenon, on the topological properties of prototype devices. }

{Previous studies considered }Fabry-Perot-induced transmission resonances in gated graphene junctions~\cite{Beenakker_Andreev_graphene_2008,CastroNeto_graphene_review_2009}, but, in contrast to that case, the physics discussed here relies on the spinor structure of the TIS. Conductance oscillations were predicted in the TI surface states subject to the exchange field in the gated region~\cite{Mondal2010,Soori2012}, but those studies assume no distinction between the interface and the surface states, and ignore the scattering at lateral junctions, which we show to be a critical consideration. Our proposal
provides a quantitative measure of symmetry-breaking effects intrinsic to the interface. We also show that the results are robust with respect to the details of the junction scattering~\cite{note}.

The rest of the paper is organized as follows. In Sec.~\ref{sec:junction} we explain our model, and the methods we use to obtain the conductance. Sec.~\ref{sec:helical} gives the simplest possible example of a non-trivial junction, that with two distinct helical Dirac cones, and elucidates the essential physics at play. The symmetry-broken case is considered in Sec.~\ref{sec:spinz}, where we focus on the information on the anisotropy of the Dirac cone that can be extracted from the conductance measurements, and its connection to the spin texture of the interface states. Sec.~\ref{sec:disc} summarizes our findings, provides the context for their interpretation, and discusses possible further advances.

\section{Lateral Junctions}
\label{sec:junction}

Bulk-boundary correspondence guarantees the existence of a localized state at the interface between a TI and a non-topological material~\cite{Rev_zhang,Rev_kane}. In the absence of time-reversal breaking perturbations,  this state is generically gapless and, at low energy, linearly dispersing with the momentum in the plane of the interface, $\bm k$. Since the band inversion, which is responsible for the topological properties of the TI, arises from the spin-orbit interaction (SOI), the resulting states also characteristically exhibit spin-momentum locking~\cite{Rev_zhang, Rev_kane, Rev_hasan}.

For prototypical TI Bi$_2$Se$_3$, the states at the surface terminations along the high symmetry directions ([111] in the rhombohedral/[001] in the hexagonal representation)
are described by the effective Dirac Hamiltonian~\citep{Zhang2009}
\begin{equation}
  H_D = v_1(\bm{\sigma}\times \bm{k})_z\,,
  \label{Ham_D}
\end{equation}
where $\bm\sigma$ is the vector of the Pauli matrices, $\widehat{\bm z}$ is the normal to the surface, $v_1$ is the effective velocity, $\bm k$ is the momentum in the $x$-$y$ plane, and we set $\hbar=1$.
The eigenstates of Eq.~\eqref{Ham_D} have isotropic dispersion, $E_{\pm} = \pm  v_1{k}$, and are helical (spin in the interface plane normal to the momentum direction), with the spinor structure
\begin{equation}
\widetilde\Psi_D (E, \bm k)=\begin{pmatrix}
i\\ \mbox{sgn}(E)e^{i\theta}\\
\end{pmatrix}\,,
\label{spinor_D}
\end{equation}
where $\theta$ the angle between ${\bm k}$ and the $x$-axis.

Once the rotation symmetry is broken either by a choice of the lattice termination~\cite{FZhang2012} or by the interface potentials~\cite{mahmoud_se_ti}, the topological state is described by a generalized form of $H_D$, namely
\begin{equation}
  H_I = \sum_{i,j}c_{ij}\sigma_i k_j\,,
\label{Ham0}
\end{equation}
{where the sum is over all the spin components, $i=x,y,z$, but the momentum is in the plane, $j=x,y$, and the coefficients $c_{ij}$ are real. The eigenstates of $H_I$ 
have anisotropic dispersion,
\begin{equation}
  E(\bm k)=\pm\sqrt{\sum_i\left(\sum_j c_{ij}k_j\right)^2}\,,
  \label{E_anis}
\end{equation}
with non-helical spin texture and spins that point out of the plane if $c_{zi}\neq 0$. It is important to note that this Hamiltonian arises from the microscopic analysis, so that the terms leading to the anisotropy are {\em linear} in $\bm k$, and the coefficients $c_{ij}$ are uniquely determined by the bulk properties and interface potentials~\cite{mahmoud_se_ti}. These potentials depend on the details of the constituent materials and growth process, and hence a priori are unknown.  Therefore, for a generic interfacial topological state, the anisotropic dispersion persists at low energies, in contrast to the anisotropies induced in the surface states by the higher order, in $\bm k$, terms proposed~\cite{Fu_hex2009}, for example, for Bi$_2$Se$_3$ and Bi$_2$Te$_3$ due to hexagonal warping of the crystal. It is also important to note that the dependence of $c_{ij}$ on the interface potentials is such that the Hamiltonian, Eq.~\eqref{Ham0}, reduced to Dirac hamiltonian, Eq.~\eqref{Ham_D}, for the surface states, i.e. when any of these potentials allowed by symmetry become large~\cite{mahmoud_se_ti}.}


{Given that there are six independent coefficients in Eq.~\eqref{Ham0}, complete characterization of a specific interface requires performing several indirect measurements, and modeling the results. A simpler, and more practical, question, is what symmetries are broken at a specific interface, and what the consequences of this symmetry breaking for the dispersion and spin-momentum locking of the topological state are. This is the task that, as we show, can be accomplished via conductance measurements on lateral junctions, and below we provide a detailed analysis of such a setup. }

The choice of a lateral junction, surface state in prototypical TIs are well established, and therefore the parameter $v_1$ in the surface Dirac Hamiltonian, Eq.~\eqref{Ham_D}, is known. As we see below, it is gating of the interface region that enables the information on the symmetry breaking in the hidden TIS to be extracted from conductance measurements. In the following we assume that the bottom surface of the TI is far enough to not hybridize with the top surface, and therefore focus our attention on the top surface near the contacts.

\begin{figure}[t]
\includegraphics[width=0.95\columnwidth]{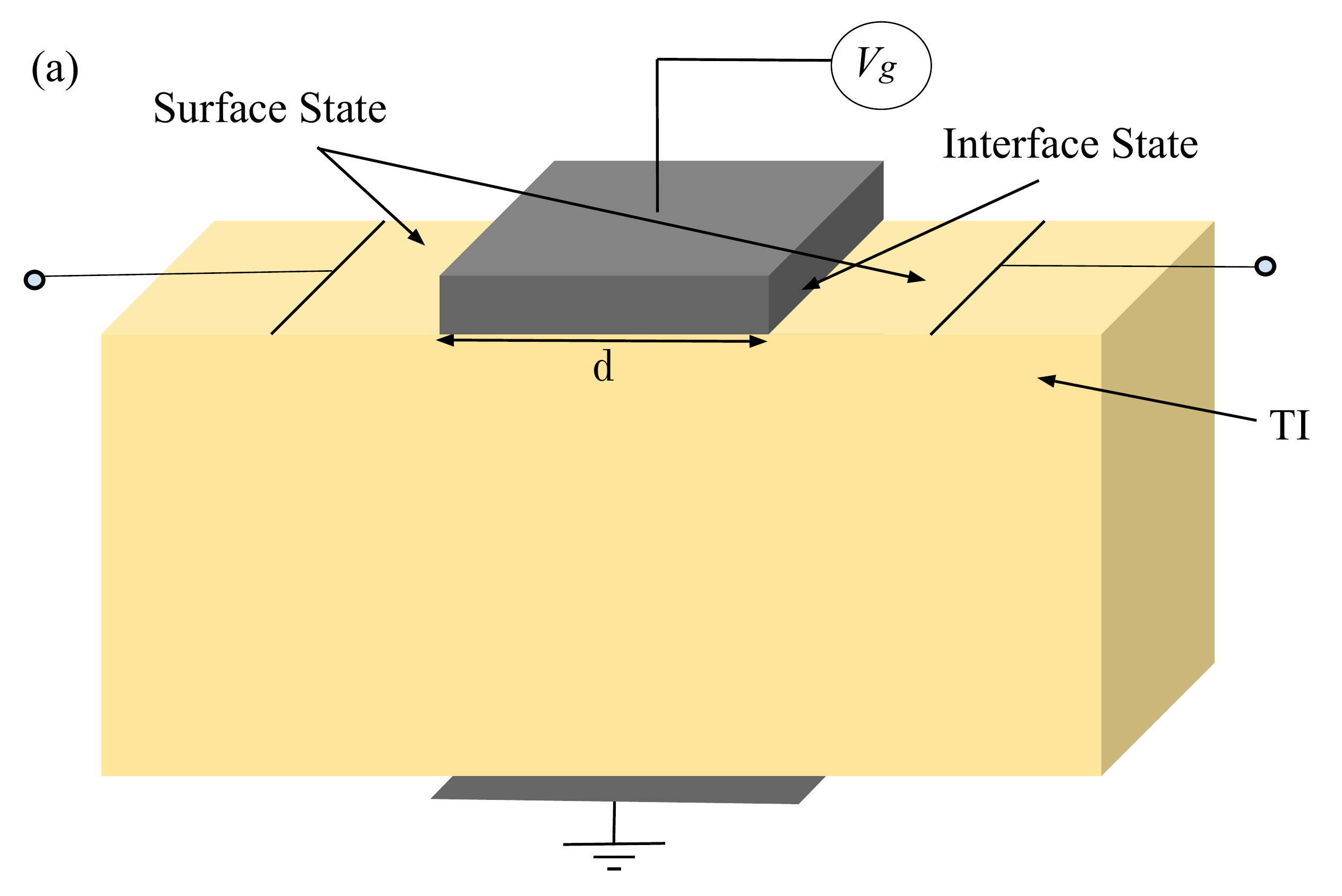}
\includegraphics[width=0.95\columnwidth]{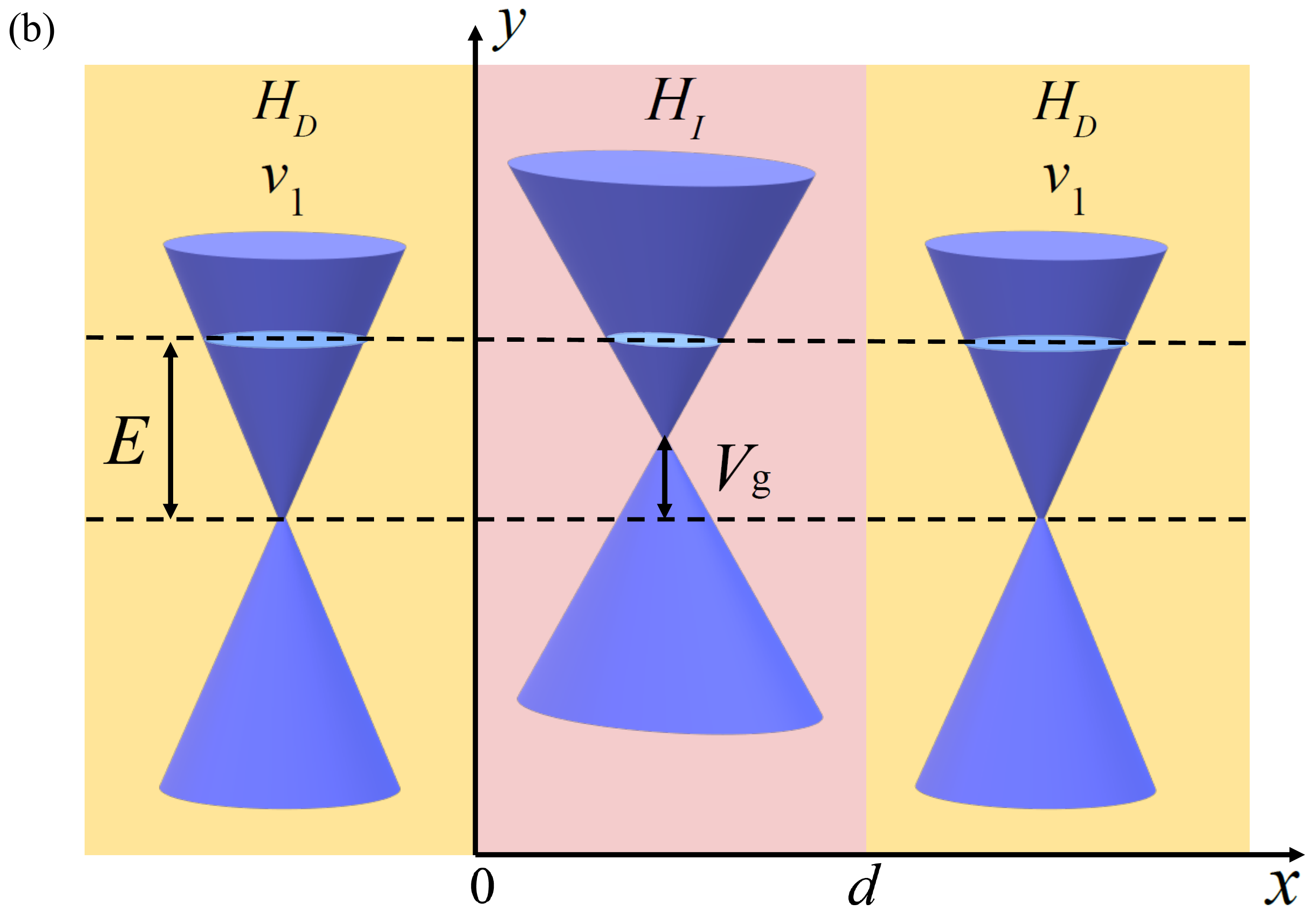}
\caption{\small Double junction setup for measuring the conductance oscillations. Top panel: schematics of the suggested measurement. 
Bottom panel: view of the top surface of the structure with the Dirac cones of the topological states, subject to the gate voltage $V_g$. $E$ is the energy of a particle relative to the Dirac point.}
\label{fig:device}
\end{figure}

The two types of Dirac dispersion, for the surface and the interface states respectively,  are shown in Fig.~\ref{fig:device}(b). We model this junction by a piecewise Hamiltonian,
\begin{equation}
  H=\begin{cases}
    H_{D} &\mbox{ for }x<0 \mbox{ and } x>d\,,\\
    H_I+V_g\openone & \mbox{ for }0<x<d\,.\\
  \end{cases}
  \label{Ham-junc}
\end{equation}
The propagating eigenstates in each region are
\begin{eqnarray}
  \Psi_D(E,x,y)&=&\widetilde\Psi_{D}(E,\bm k) \exp(i\bm k\cdot \bm r)\,,
  \\
  \Psi_I(E, x,y)&=&\widetilde\Psi_{I}(E,\bm k^\prime)\exp(i\bm k^\prime \cdot \bm r)\,.
\end{eqnarray}
Here $\widetilde\Psi_{D}(E,\bm k)$ is the spinor of the helical Dirac state, Eq.~\eqref{spinor_D}, and $\widetilde\Psi_{I}(E,\bm k)$ is the corresponding spinor for $H_I$, which depends on the choice of the coefficients $c_{ij}$, and will be given below for the specific cases we consider. The solutions of the eigenvalue equation $H\Psi=E\Psi$ are constructed by matching these solutions at the boundaries.


The energy and the momentum along the junction ($y$-axis in Fig.~\ref{fig:device}(b)) is conserved, $k_y(E)=k^\prime_y(E)$. Since the Hamiltonian is linear in $\bm k$, and the momentum along the junction ($y$-axis in Fig.~\ref{fig:device}(b)) is conserved, the eigenvalue differential equation along the normal ($x$ direction) is of first order. Hence the wave functions need not be continuous~\cite{falko,basko,akhmerov}, but, instead, at $x_0=0,d$ have to satisfy 
\begin{equation}
  \Psi (x_0^-,y)=M\Psi(x_0^+,y)\,,
  \label{M_gen}
\end{equation}
where $x_0^\pm$ denotes the side of the junction. Since spin is conserved in scattering on non-magnetic potentials, the form of the matrix $M$, which connects the spinors on both sides of the junction, relates the spin content of the transmitted, incoming, and reflected states.
The matrix $M$ has to preserve the time-reversal invariance of the Hamiltonian, i.e. $[i\sigma_y\mathcal{C},M]=0$, where $\mathcal{C}$ denotes complex conjugation. It also has to conserve the particle current normal to the junction.

The particle current operator in each region is given $j_n=\delta H/\delta k_n$. Therefore, for a piecewise defined Hamiltonian, the current operator is also defined in each region, and the current conservation at each junction demands that
\begin{eqnarray}
  &&\braket{\Psi(x_0^+,y)|j_n^+|\Psi(x_0^+,y)} = \braket{\Psi (x_0^-,y)|j_n^-|\Psi (x_0^-,y)}
  \nonumber
  \\ &&\qquad\qquad = \braket{\Psi(x_0^+,y)|M^\dagger j_n^- M|\Psi(x_0^+,y)}\,.
\end{eqnarray}
It follows that the boundary matrix $M$ must satisfy
\begin{equation}
  M^\dagger j_n^- M = j_n^+\,.
  \label{M_cur}
\end{equation}
For linear in $\bm k$ hamiltonians  $j_n$  is momentum-independent. Under these conditions Eq.~\eqref{M_cur} defines a single parameter family of matrices~\cite{Sen2012,Isaev2015,Tanhayi2016}, $M(\alpha)$.  The physical origin of this degree of freedom is that in the microscopic formulation time-reversal invariance allows spin-independent potential barrier at the junction. The height of this barrier (apriori unknown and difficult to control in experiment) is encoded in the parameter $\alpha$. The exact relationship between the two can be derived~\cite{EThareja1}, but is not important for the discussion below. Assuming that we cannot experimentally control the value of $\alpha$ at each junction implies that we need to show that our predictions are insensitive to it.


Below, for several choices of the symmetry of $H_I$, we determine the matrix $M$, and use it to compute transmission ($\mathcal{T}$) and scattering ($\mathcal{S}$) matrices, see Appendix. This yields the transmission coefficient, $T (E,\theta)$ as a function of the energy, $E$, and the incident angle, $\theta$, for different gate voltages, $V_g$, and the boundary coefficients $\alpha,\beta$ at each junction. For a single junction the transmission is dominated by Klein tunneling at normal incidence, $T(E,\theta=0)=1$, irrespective of the choice of $c_{ij}$, $\alpha,\beta$.  The anisotropy of the Dirac states, and details of the junction potentials appear as very weak features for nearly grazing angles, and therefore are not clearly manifested in the conductance. These features, however, are brought to the fore by the Fabri-Perot oscillations in the double junction setup of Fig.~\ref{fig:device}(a). We use the transmission coefficients for the double junction to evaluate the low temperature conductance using Landauer formalism,
\begin{equation}
  \frac{G(V_g)}{G_0} = \frac{1}{2}\int_{-\pi/2}^{\pi/2} T(\mu,\theta) \cos{\theta} d\theta\,,
  \label{G_gen}
\end{equation}
where $G_0 = \frac{e^2}{h\pi}Wk_F$, $W$ the width of the junction, $\mu$ is the chemical potential, and the Fermi momentum is determined from the dispersion of the surface states, $\mu=v_1k_F$.

Gating the interface state by the gate voltage $V_g$ changes the size of the Fermi surface in the interface region, see Fig.~\ref{fig:device}(b). Together with the momentum conservation along the junction, this enables directional-dependent probing of the Dirac cone. The key to understanding the physics behind this directional dependence is to recognize that there are two distinct regimes of the behavior of the conductance as a function of $V_g$. These regimes are easily understood from the analysis of the simplest cases, presented in next section.

It is appropriate to comment on the potential experimentally relevant aspects of the physics that are left out of this model. First, we consider only the surface and interface states, and neglect the contribution of the bulk bands. Since in experiment the chemical potential can be tuned by alloying ~\cite{Zhang2011,Arakane2012}, we believe that this regime is experimentally achievable. In more general terms the applications of the TIs mostly rely on eliminating, or substantially reducing, the bulk contributions, and therefore we follow a well-trodden path in making this assumption. Second, in some TIs a 2DEG forms after some time in the vacuum chamber~\cite{Bianchi2010}. Potentially this can be avoided by capping the TI with a very large gap semiconductor, when the distortion of the surface states is minimal~\cite{mahmoud_se_ti}. Moreover, this 2DEG is formed at energies of the order of the bulk gap, and therefore tuning the chemical potential close to the Dirac point and using small gate voltages ensures that it does not poison the conductance measurements. The same small values of the gate voltage relative to the bulk gap value ensure that we do not need to consider hexagonal distortion that is cubic in the distance to the Dirac point in the momentum space, and therefore is not relevant at low energies. In all of the following we restrict ourselves to parameters satisfying these constraints.

\section{Helical Interface State}
\label{sec:helical}

\subsection{Conductance of gated junctions and the Dirac velocity.}

\begin{figure}[t]
\includegraphics[width=\columnwidth]{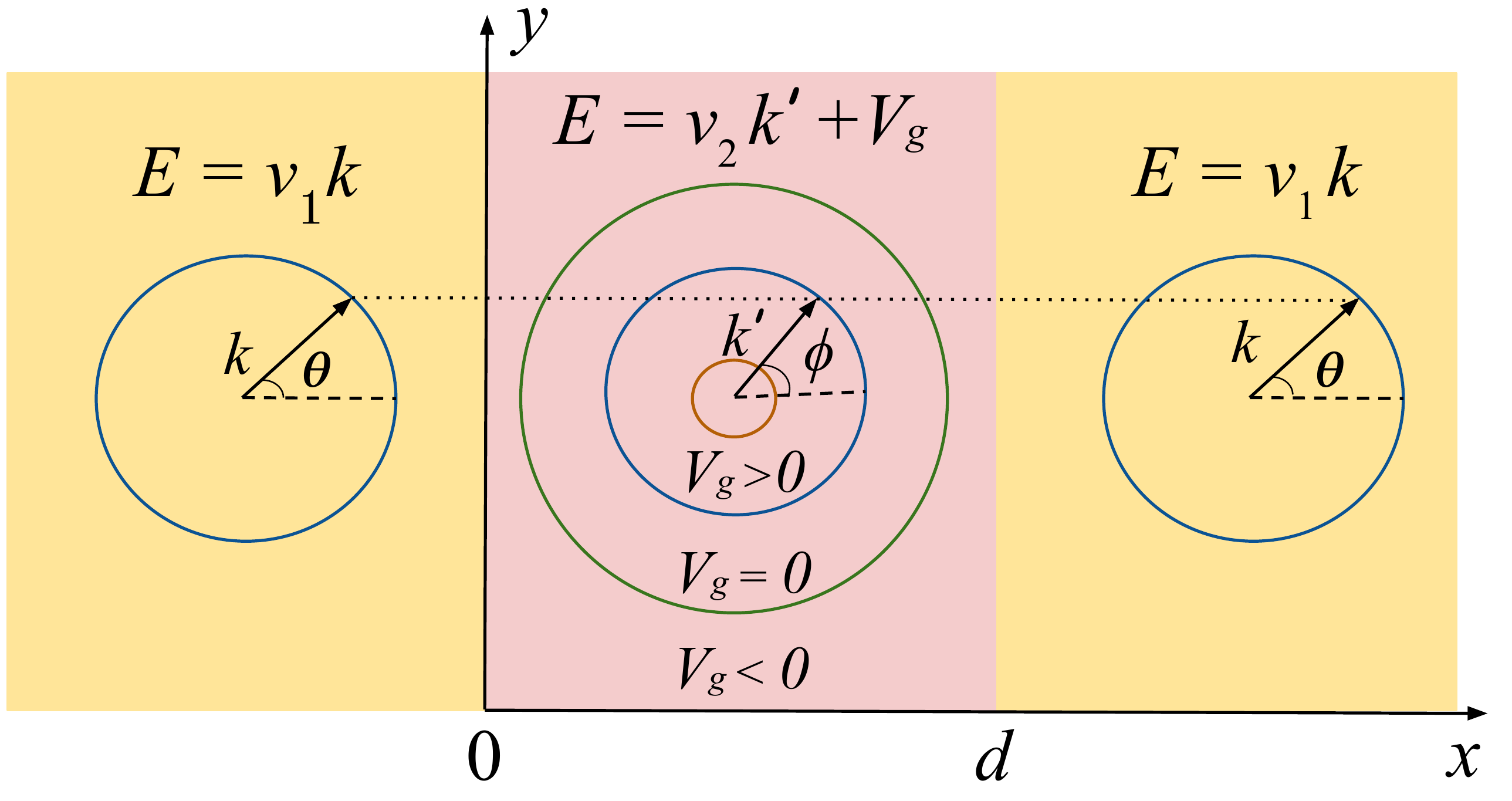}
\caption{Top view of a double junction with helical interface state. Gating changes the size of the constant energy surface in the interface region. Conservation of the momentum along the junction (dashed line) implies that for $V_g>0$ only electrons occupying the surface states at near normal incidence on the left side enter the interface region, where they continue to propagate covering all incidence angles for the right junction. In contrast, for $V_g<0$, all the electrons from the surface states on the left can enter the interface region, but then come at near normal incidence to the right junction. }
\label{fig:helic}
\end{figure}

Most of the relevant physics can be inferred from considering the simplest possible case, where the interface state is also helical, $H_I = v_2(\bm{\sigma}\times \bm{k})_z$, as illustrated in Fig.~\ref{fig:helic}.
One of the simplifications is that the matrix form of the particle current operator in the direction normal to the junction is the same in all regions defined by the Hamiltonian, Eq.~\eqref{Ham-junc}, namely $j_x^\pm=-v_{1,2}\sigma_y$. Substituting in Eq.~\eqref{M_cur} yields
\begin{eqnarray}
  M_L&=&\sqrt{\frac{v_2}{v_1}}\exp[i\alpha\sigma_y]\,,
  \\
  M_R&=&\sqrt{\frac{v_1}{v_2}}\exp[i\beta\sigma_y]
\end{eqnarray}
for the junctions at $x=0$ and $x=d$ respectively, with $\alpha,\beta$ arbitrary real parameters. The form of the boundary matrix is easy to understand if we realize that the rotations in the spin space around the $y$-axis leave the current operator $j_x$ invariant. The ratio of the velocities ensures conservation of the probability current.

The role of gating is qualitatively clear from Fig.~\ref{fig:helic}, where we graphically show the consequences of the conservation of the energy, $E=v_1k=v_2k^\prime+V_g$ (note that we plot only $V_g<\mu$, since, in the low energy theory, the physical picture is symmetric with respect to $V_g=\mu$, and we discuss corrections to this picture separately below), and the momentum along the junction,
\begin{equation}
  k\sin\theta=k^\prime\sin\phi\,,
  \label{k_y_cons}
\end{equation}
where $\theta$ and $\phi$ are the directions of propagation of the quasiparticles in the surface and interface regions respectively. At low temperatures, $kT\ll\mu$,   we can set $E=\mu$ for the quasiparticles contributing to the conductance without loss of generality. For sizeable positive bias, $V_g>0$, only the particles at near-normal incidence from the region $x<0$ may enter the interface region. The transmission for near-normal incidence is close to unity, but the density of states in the interface region is low due to small size of the Fermi surface, and is the main limiting factor for transport. Consequently, we expect the conductance to be essentially linear in the gate voltage, $|\mu-V_g|$, in the vicinity of  $G(V_g=\mu)=0$ (quasiparticle energy in the interface region tuned to the Dirac point), with additional, weaker, variations due to the matching conditions at the second junction, $x=d$.

For sufficiently negative bias, $V_g<0$, the momentum conservation dictates that all the incident quasiparticles from the surface states travel nearly along the $x$-axis in the interface region. Consequently we expect Fabri-Perot oscillations with the phase acquired in this region, with maxima at
\begin{equation}
  k^\prime d\equiv\frac{\mu-V_g}{v_2} d=\pi n\,, \mbox{ with } n=0,1,\ldots\,,
\end{equation}
corresponding to the periodicity with the gate voltage,
\begin{equation}
  \Delta V_g\approx \frac{\pi v_2}{d}\,.
  \label{V_g_period}
\end{equation}
Note that, because of Klein tunneling, the quasiparticles at {\em exactly} normal incidence do not contribute to the oscillations. Therefore, the overall transmission is high (conductance $G(V_g)/G_0\sim 1$), and we expect the amplitude of the oscillations to be a fraction of the total conductance, in contrast to the usual Fabri-Perot interferometer. We see below that this fraction, however, is substantial under realistic assumptions.

The crossover between the two regimes occurs when the two Fermi surfaces are of the same size, at the gate voltages
\begin{equation}
  V_k^{(\pm)} =\mu(1\pm v_2/v_1)\,,
  \label{kink_V}
\end{equation}
which we label ``kink'' voltages hereafter. For $V_k^{(-)} <V_g< V_k^{(+)}$ the density of states effects dominate, and the conductance is suppressed. Outside of this range the double junction is highly transparent, and the oscillations due to the phase acquired in the interface region are clearly observable.

Detailed calculations confirm these expectations, and allow the analysis of their dependence on the boundary angles $\alpha$ and $\beta$.  Appendix~\ref{sec:app_hel} gives the details of the derivation of the transmission coefficient by matching the wave functions at both linear junctions, and yields
\begin{widetext}
\begin{equation}
T 
= \frac{\cos^2\theta  \cos^2 \phi }{{|\cos \phi  \cos \zeta (\cos \theta  \cos \eta_+ -i \sin \eta_+ )
 +\sin \zeta [i \sin \theta  \sin \phi  \cos \eta_- }-s(\cos \theta  \sin \eta_+ +i \cos \eta_+ )]|^2}\,,
\label{trans_gen}
\end{equation}
\end{widetext}
where we denoted the phase accumulated while traversing the interface region $\zeta=k^\prime d\cos\phi$, and introduced $\eta_\pm = \alpha \pm \beta$, and $s = \mathrm{sgn}(\mu-V_g)$. This expression, together with Eq.~\eqref{G_gen}, is used to evaluate the conductance for specific parameter values.

We set the Dirac velocity of the surface state to its experimental value~\cite{Kuroda2010,Chang2015} for Bi$_2$Se$_3$,~\cite{Kuroda2010,Chang2015} $v_1=3.3 \hspace{1mm} \mathrm{eV \cdot\AA}$. We use $\mu = 0.02$ eV in most of our calculations, so that we stay close to the Dirac point, and can ignore the hexagonal warping, bulk band transport, and possible formation of the 2DEG at the interface. This choice gives $k_F=6.6\times 10^{-3}\ $\AA$^{-1}$. Realistic device sizes then imply $k_Fd\geq 1$. In most of our calculations we set $d=40$ nm ($k_Fd\approx 2.4$), which allows coherent quasiparticle propagation across the gated region according to the values of the mean free path reported for $\mathrm{Bi_2Te_xSe_{3-x}}$ and Bi$_2$Se$_3$~\cite{Dufoul2017,Kamboj2017,Chen2013}($20-120$ nm), and especially for $\mu$ close
to the Dirac point~\cite{Kellner2015}, but also consider other values of $d$ for the situation where this is relevant, see below.

\begin{figure}[b]
\includegraphics[width=0.5\textwidth]{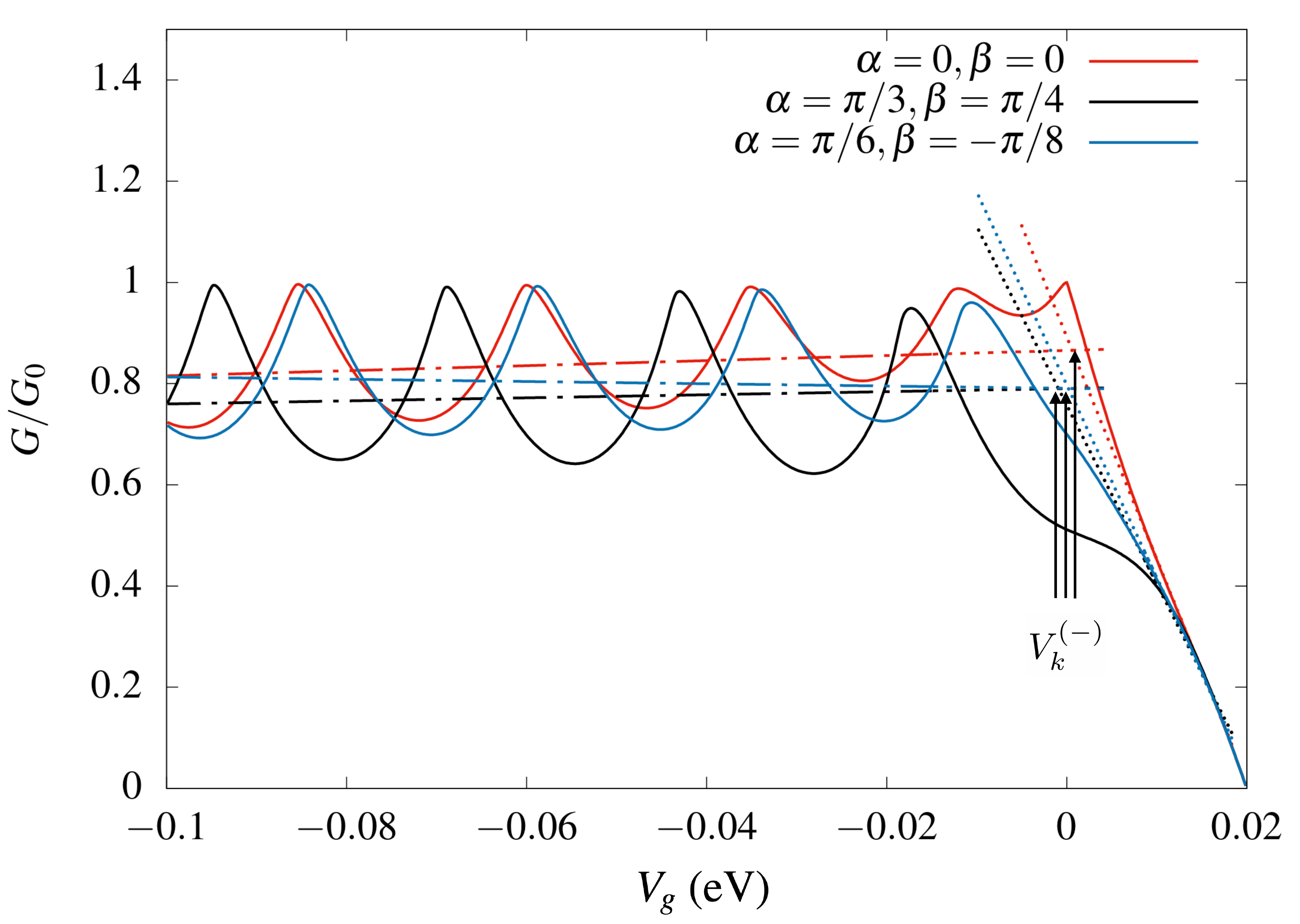}
\caption{Conductance oscillations for  $v_1 = v_2=3.3 \hspace{1mm} \mathrm{eV \cdot\AA}$, with our choice of $\mu=0.02$ eV, $d=40$ nm. Note that the kink voltage does not depend on the scattering parameters at the junction, see text.}
\label{fig:conduct_eq}
\end{figure}

It is important to understand that the oscillations are present even if the Dirac velocities in the two regions are identical, $v_1=v_2$, as shown in Fig.~\ref{fig:conduct_eq}. The gated region creates mismatch of the Fermi surfaces, and the conductance shows the two regimes discussed above, namely the linear decrease near $V_g=\mu$, and the oscillatory high conductance region away from that value. Importantly, Fig.~\ref{fig:conduct_eq} shows that the kink voltage, in this case $V_k^{(-)}=0$ (cf. Eq.~\eqref{kink_V}), is insensitive to the junction scattering parameters $\alpha,\beta$. {For practical purposes, we define this voltage as the intersect between linear fit to $G(V_g)$ in two regimes. First, at greater $V_g$, we average the conductance over one oscillation period, and perform a linear fit for the resulting values. Second, we extend the linear $G(V_g)$ near $V_g=\mu$. These lines are shown in Fig.~\ref{fig:conduct_eq}, and their intercept defines $V_k^{(-)}$.} Even though the slope near $V_g=\mu$ varies by about 20\% as $\alpha,\beta$ vary, the intercept only changes by about 5\% across those values. We find the same behavior for $v_1\neq v_2$; Moreover, we find that the difference between the kink voltages determined from the intercept for different $\alpha,\beta$, and the value given by Eq.~\eqref{kink_V} decreases with the increased mismatch of the Dirac velocities.

\begin{figure}[t]
\includegraphics[width=\columnwidth]{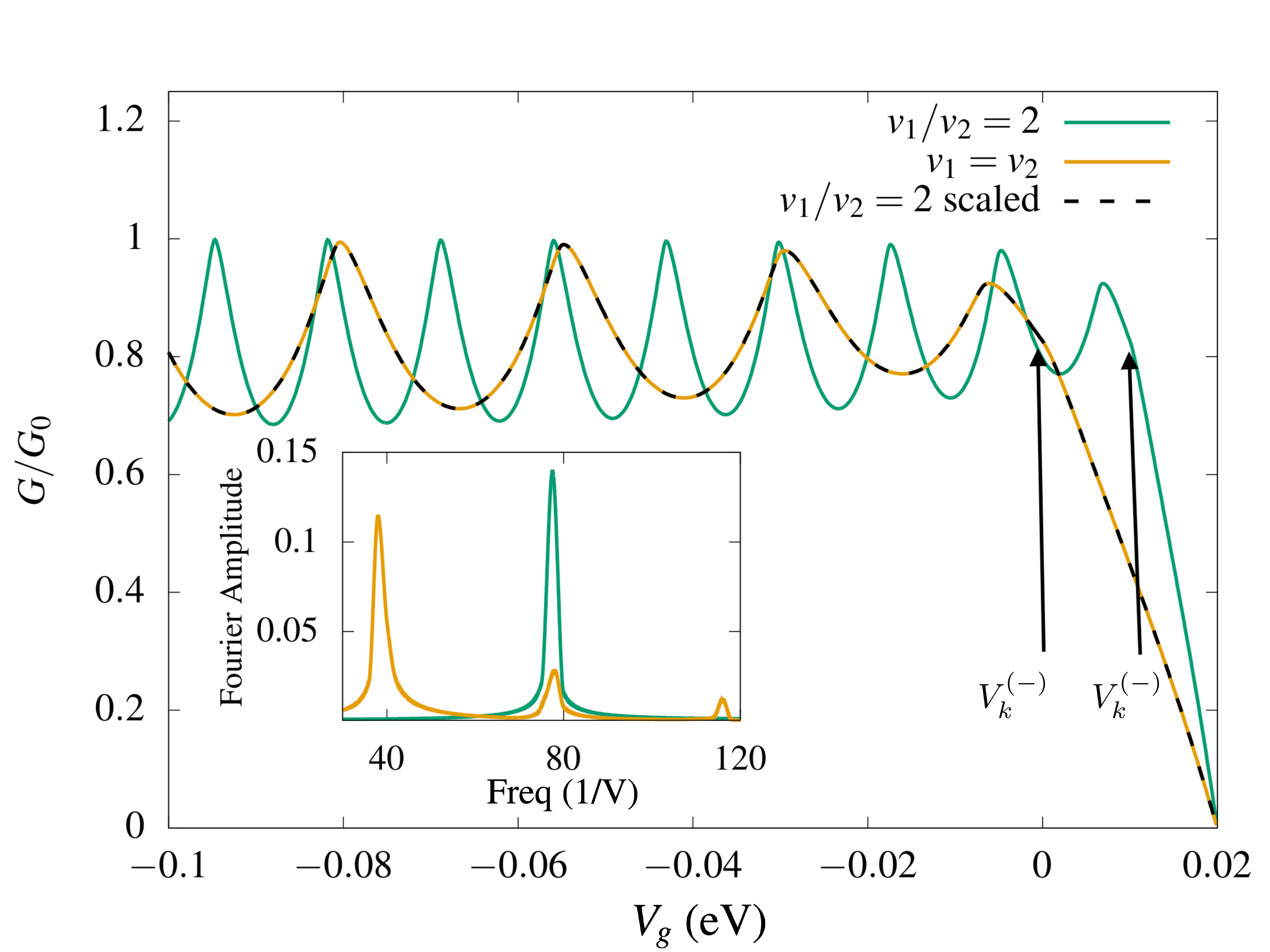}
\caption{\small Conductance as a function of gate voltage for the junction of helical Dirac states, with $\mu=0.02$ eV,$\alpha=0, \beta=\pi/6$.  Main panel: Conductance $G(V_g)$ shows both the quasi-linear regime around $V_g=\mu$ and oscillatory behavior away from it, separated by the kink voltage, $V_k^{(-)}$, indicated by arrows for the two different choices of the Dirac velocities.  Rescaling $V_g$ by the ratio $v_2/v_1$ makes the plots identical. Inset: Fourier transform of oscillatory part of the conductance, see text. } 
\label{fig:G_osc}
\end{figure}

\begin{figure}[ht]
\includegraphics[width=0.5\textwidth]{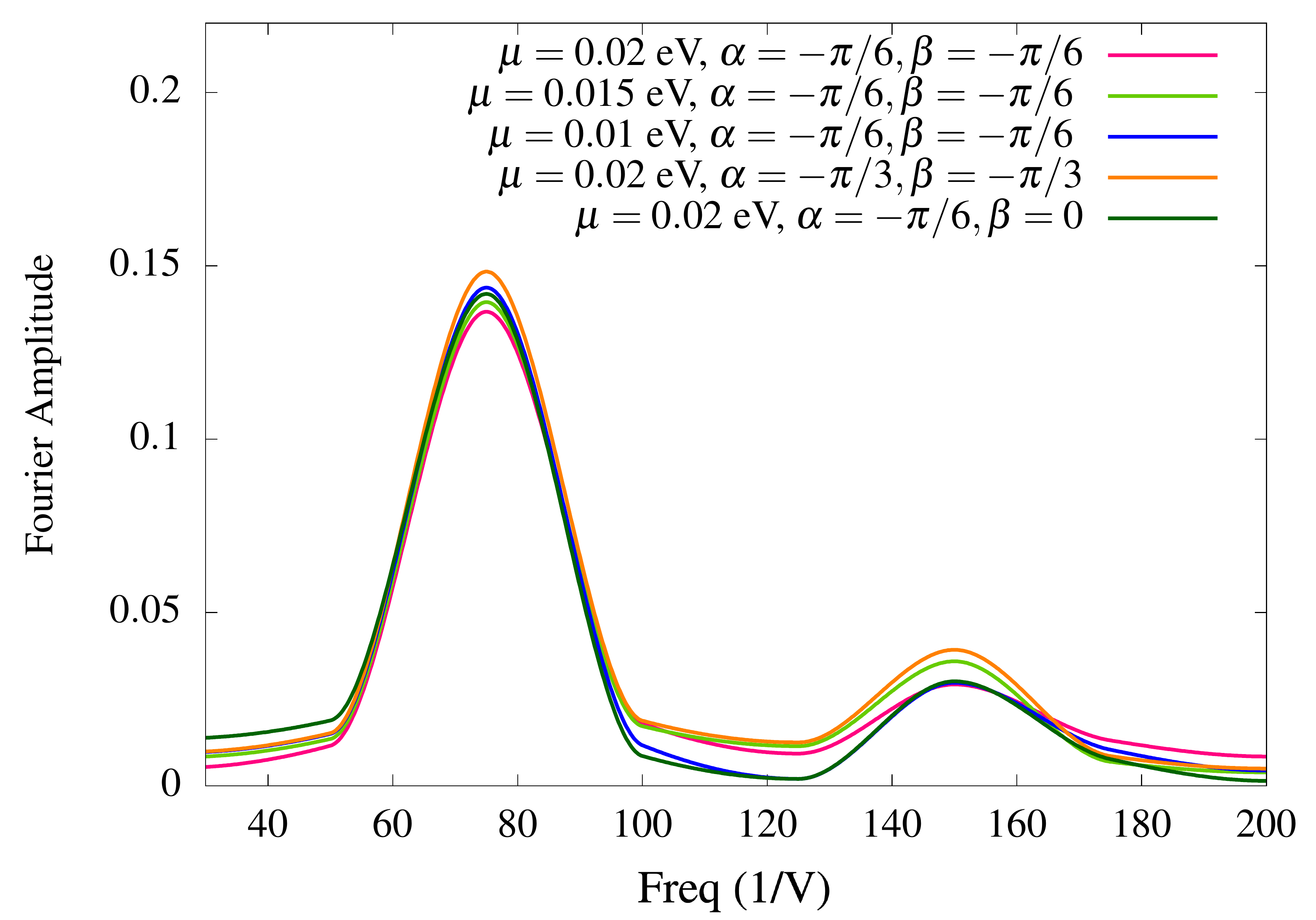}
\caption{Lead frequency of conductance oscillations for different values of $\alpha$, $\beta$ and $\mu$. Values of other parameters used: $v_1 = 3.3$ eV {\AA} and $v_2 = 1.65$ eV {\AA}}
\label{fig:freq}
\end{figure}

It is also clear from Fig.~\ref{fig:conduct_eq} that junction scattering changes the phase of the oscillations at $-V_g\geq \mu$, but leaves the period unchanged, and we now show this in more detail by comparing the results for $v_2=v_1$ and $v_2=v_1/2$ in Fig.~\ref{fig:G_osc} for a select (non-trivial choice of junction scattering. Arrows indicate the kink voltages determined from the intercept as described above, and correspond very well to the values of $0$ and $0.01$eV respectively. Moreover, the plots for the two cases coincide exactly after rescaling the gate voltage by the ratio $v_2/v_1$, verifying that the Dirac velocity is the sole control parameter for both the kink voltage and the oscillation frequency in the double junction. Inset shows the Fourier transform of the oscillatory part of the conductance, and the main peaks of the Fourier amplitude correspond very well to the frequencies $f_l=38.6$eV$^{-1}$ and $77.2$eV$^{-1}$ for $v_2=v_1$ and $v_2=v_1/2$ respectively, obtained from Eq.~\eqref{V_g_period}. We also verified that these frequencies are insensitive to the choice of the scattering parameters, or, equally important, to the exact choice of the chemical potential, $\mu$, as shown in Fig.~\ref{fig:freq}.

Therefore the profile of the conductance in a gated double junction yields, for fully symmetric surface and interface states,  two distinct methods for evaluating the Dirac velocity in the interface region: from the kink voltage, $V_k^{(-)}=\mu(1-v_2/v_1)$, and from the  oscillation period, $\Delta V_g\approx \pi v_2/d$. In the following section we show that, for anisotropic, symmetry-broken, interface states, the two methods give the values for the Dirac velocity in the two orthogonal directions, parallel and normal to the junction, and therefore can be used to quantitatively determine the anisotropy of the Dirac dispersion. Before we do that it is helpful to understand better the origin of the oscillations.

\subsection{Transmission probability and conductance oscillations.}

We focus first on the negative gate voltages: while the simple low energy theory is symmetric with respect to the Dirac point, in many topological insulators of the Bi$_2$X$_3$ family the Dirac point is close to the top of the valence band, making this a natural choice. From Eq.~\eqref{k_y_cons} the direction along which a quasiparticle travels in the interface region is
\begin{equation}
  \sin\phi=\frac{k}{k^\prime}\sin\theta=\frac{\mu-V_k^{(-)}}{\mu-V_g}\sin\theta\,.
\end{equation}
Therefore for sufficiently large bias, compared to the kink voltage, when $\mu-V_g\gg |\mu- V_k^{(-)}|$, we have $\sin\phi\ll\sin\theta$, and quasiparticles in the gated region move nearly normal to the interface, see Fig.~\ref{fig:helic}. If we set $\cos\phi=0$ in the expression for the transmission coefficient, Eq.~\eqref{trans_gen}, we obtain
\begin{equation}
  T\approx \frac{\cos^2\theta}{\cos^2\theta  \cos^2 \phi+\sin^2\widetilde\zeta \sin^2\theta}\,,
  \label{trans_simple}
\end{equation}
where $\widetilde\zeta=\zeta_0+\alpha+\beta$, and  $\zeta_0=k^\prime d=k_f d(\mu-V_g)/(\mu-V_k^{(-)})$. This yields an
 approximate analytical expression for the conductance
\begin{equation}
  \frac{G(V_g)}{G_0}\approx \int_0^1\frac{dx (1-x^2)}{1-x^2\cos^2\widetilde\zeta}=F[\cos\widetilde\zeta]\,,
  \label{eq:G_approx}
\end{equation}
with
\begin{equation}
  F[x]=\frac{1}{x^2}+\frac{x^2-1}{2x^3}\ln\frac{1+x}{1-x}\,.
\end{equation}

\begin{figure}[t]
\subfloat{\includegraphics[width=0.5\textwidth]{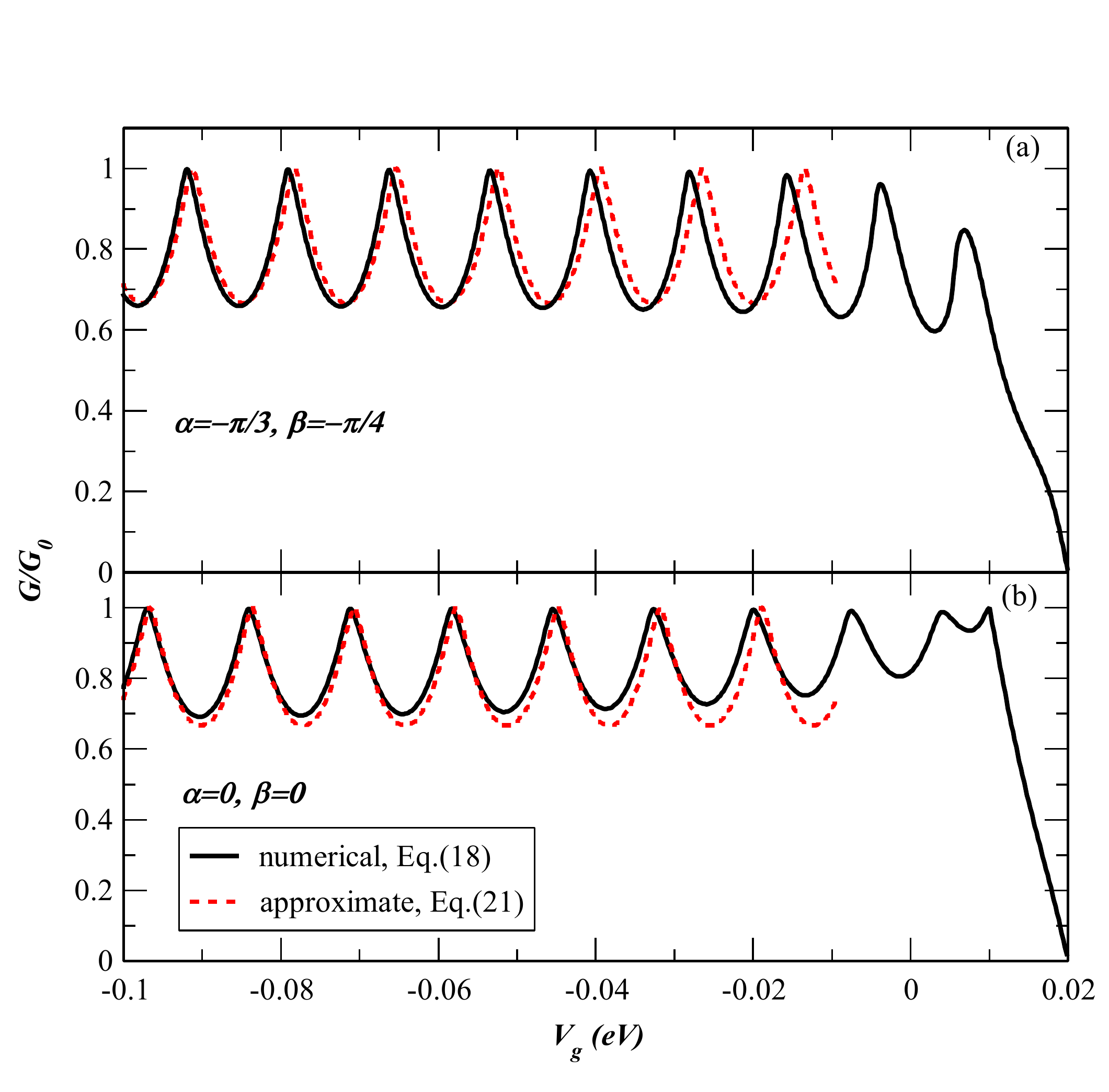}}
\caption{ {Comparison of numerical results with approximate conductance from Eq.~\eqref{eq:G_approx}. Panels (a) and (b) differ by the choice of $\alpha,\beta$, but both are for $\mu = 0.02$ eV, $v_1 = 3.3$ eV $\mathrm{\AA}$  and $v_1 = 1.65$ eV $\mathrm{\AA}$.}}
\label{fig:model_comparison}
\end{figure}

Fig.~\ref{fig:model_comparison} shows that this expressions provides an excellent agreement with our results. Eq.~\eqref{eq:G_approx} also explains why inclusion of the boundary scattering gives simply a phase shift at large gate voltages, and gives a seemingly universal amplitude of the oscillations, since the maxima and minima of the function $F$ are $F[1]=1$ and $F[0]=2/3$ respectively.

The evolution of the oscillations with the width of the gated region is more complicated, however. For large $V_g$ assuming normal travel in the gated region is justified everywhere in Eq.~\eqref{trans_gen} {\em except} in the Fabri-Perot phase, $\zeta=\zeta_0\cos\phi$, where the small variations in the angle $\phi$ are multiplied by a potentially large factor $\zeta_0$, leading to substantial variations in the contributions to the conductance. Requiring $\zeta_0(1-\cos\phi)\ll \pi$ yields a much more restrictive condition on the applicability of Eq.~\eqref{eq:G_approx}, namely
\begin{equation}
  \frac{k_F d}{\pi} \frac{\mu-V_k^{(-)}}{\mu-V_g}\ll 1\,.
  \label{large_kfd}
\end{equation}

\begin{figure}[t]
\includegraphics[width=\columnwidth]{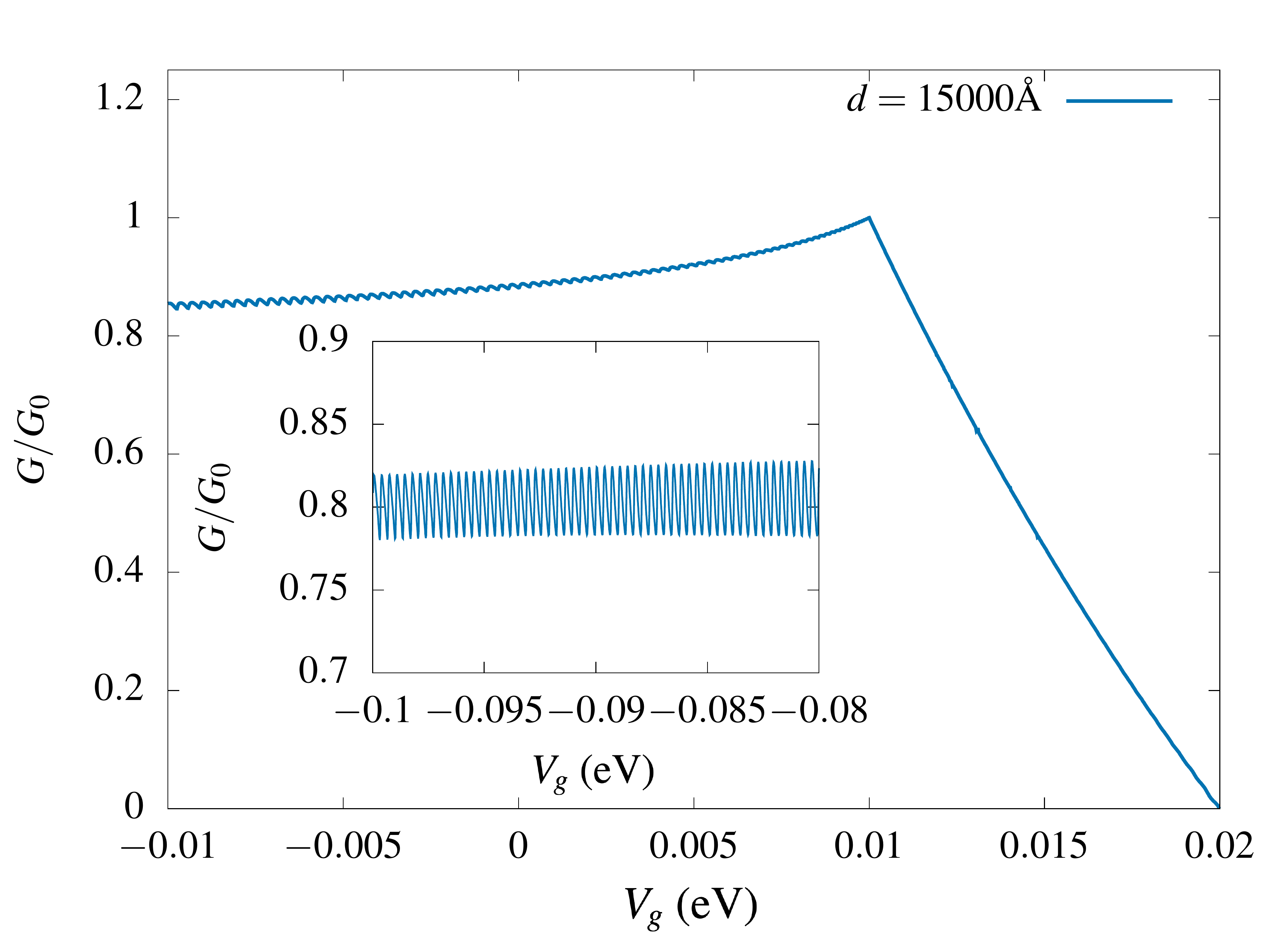}
\caption{  {Small amplitude conductance oscillations as a function of gate voltage in the large $k_Fd$ limit. Here $\mu=0.02$ eV,$\alpha=\beta=0$, $v_1=3.3$ eV {\AA} and $v_2=1.65$ eV {\AA} . Inset: At greater gate voltages the conductance, $G/G_0$, is close to the value of 0.79 obtained analytically in Eq.~\eqref{G_large_d}.  }}
\label{fig:G_large_d}
\end{figure}

In the opposite regime of large $d$, the Fabri-Perot phase oscillates rapidly relative to the variation of the incidence angle, $\sin\zeta\approx\sin(\phi_0+a\sin^2\theta)$, where $\phi_0,a\gg 1$.  Therefore we can obtain the approximate expression for the conductance by first averaging over a period of the Fabri-Perot oscillations in Eq.~\eqref{eq:G_approx}, i.e.
\begin{eqnarray}
  \frac{G(V_g)}{G_0}&\approx& \int_0^1\frac{dx (1-x^2)}{1-x^2\cos^2(\phi_0+a\sin^2\theta)}
  \label{G_large_d}
  \\
  \nonumber
  &\approx&\int_0^1 (1-x^2) dx\left[\frac{1}{\pi}\int_0^\pi \frac{dz}{1-x^2\cos^2z}\right]=\frac{\pi}{4}\,.
\end{eqnarray}
This value of the conductance, $G/G_0\approx 0.79$ is in a very good agreement with the numerical results presented in Fig.~\ref{fig:G_large_d}. As expected, the oscillation amplitude is significantly reduced in this regime.

\begin{figure}[t]
  \includegraphics[width=\columnwidth]{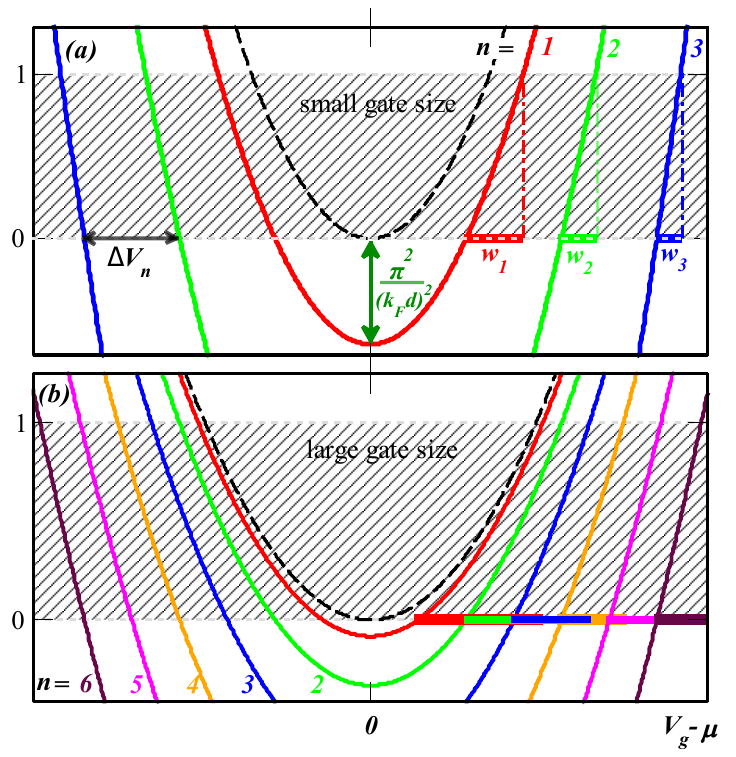}
  \caption{\small Location of the transmission maxima. Panels (a) and (b) show parabolas in the right-hand side of Eq.~\eqref{theta_n} for large and small gate length, $d$. The function must be in the $[0,1]$ interval (shaded) to have a solution for $\theta_n$. The corresponding ranges ($w_n$, shown by thick horizontal lines) of gate voltage $V_g$ for different orders, $n$, do not (do) overlap for small (large) values of $d$. $\Delta V_n$ is the interval between the appearance of two successive (generally aperiodic and non-sinusoidal, see text for details) oscillations.} 
  \label{Fig:parabola}
\end{figure}

Qualitatively, if $k^\prime d\gg 1$, the rapid oscillations of the Fabri-Perot phase with the incidence angle, $\theta$, in Eq.~\eqref{trans_simple}, lead to rapid variations of the transmission probability with $\theta$, with dominant contributions coming from the ``angles of perfect transmission'', $T=1$, which
occur: (a) for normal incidence, $\theta=\phi=0$ (Klein tunneling); (b) when $\zeta=k'd\cos{\phi}=n\pi$, corresponding to the incidence angles
\begin{equation}
  \sin^2\theta_n(\mu)=
  \left[\left(\frac{\mu-V_g}{v_2 k_F}\right)^2-\left(\frac{n\pi}{k_F d}\right)^2\right]\in\ [0,1]\,.\label{theta_n}
\end{equation}
The number of such angles for a gives voltage is determined by
\begin{equation}
  \frac{\mu-V_g}{\mu-V_k^{(-)}}\geq \frac{\pi n}{k_F d}\geq \sqrt{\left[\frac{\mu-V_g}{\mu-V_k^{(-)}}\right]^2-1}\,.
\end{equation}
It is obvious that the allowed values of $n$ here strongly depend on the magnitude of $k_Fd$. Fig.~\ref{Fig:parabola} shows a graphical solution of Eq.~\eqref{theta_n} for the cases of small ($k_F d\ll 1$) and large ($k_F d\gg 1$) gated region length. In the former case, Fig.~\ref{Fig:parabola}(a), the perfect transmission maxima only occur in narrow ranges of the gate voltages, which become more and more widely separated as $V_g$ increases. For a long gate there are sometimes several such maxima at gate voltages close to the chemical potential, see Fig.~\ref{Fig:parabola}(b). In principle, as $V_g$ increases, at large $n$ we cross over to the well-separated maxima regime akin to that of Fig.~\ref{Fig:parabola}(a) since the displacement of parabolas varies as $n^2$, but whether this regime is reached for experimentally relevant parameter values depends on the specifics of the material.

\begin{figure}[t]
\subfloat{\includegraphics[width=0.235\textwidth]{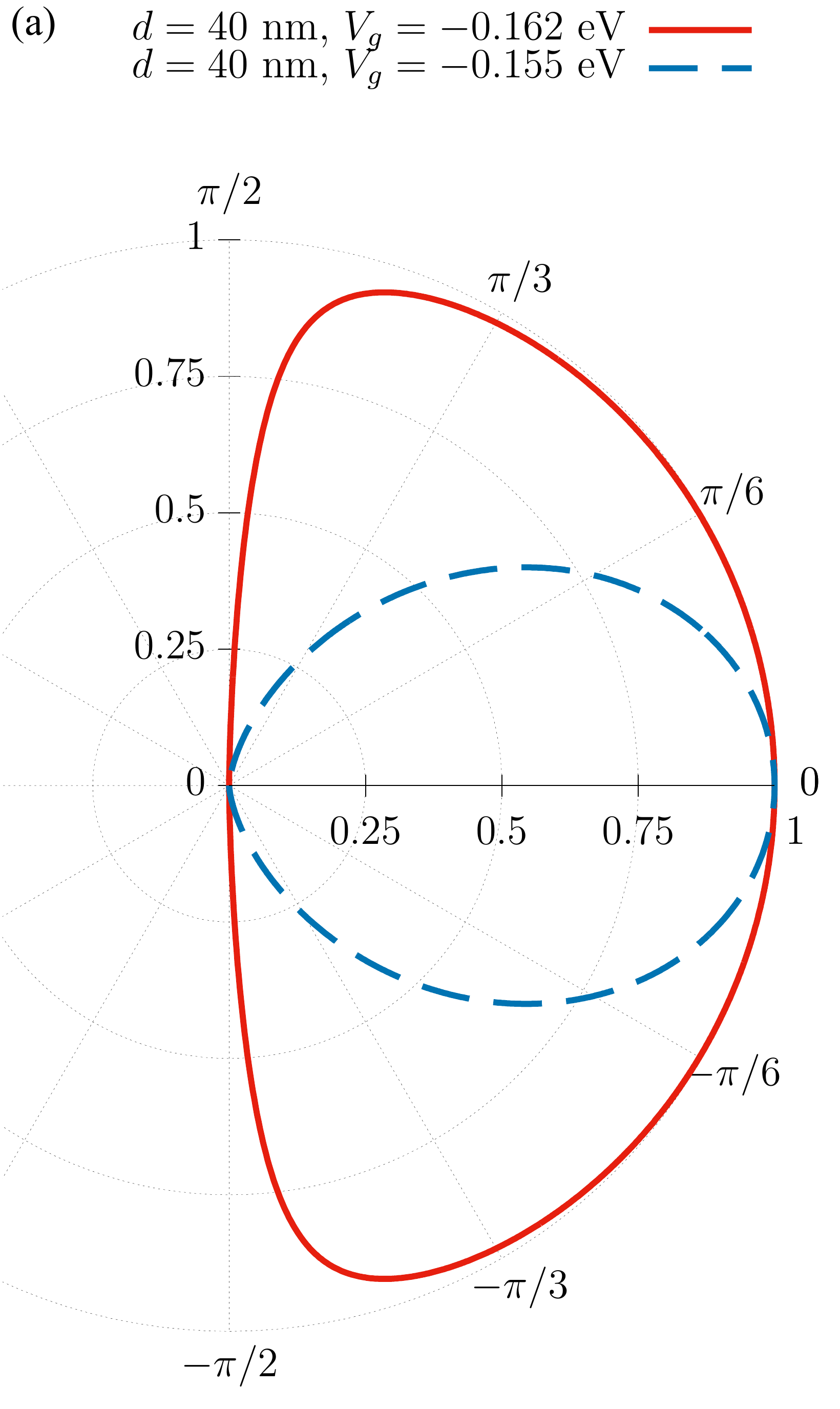}}
\hspace{0.1mm}
\subfloat{\includegraphics[width=0.23\textwidth]{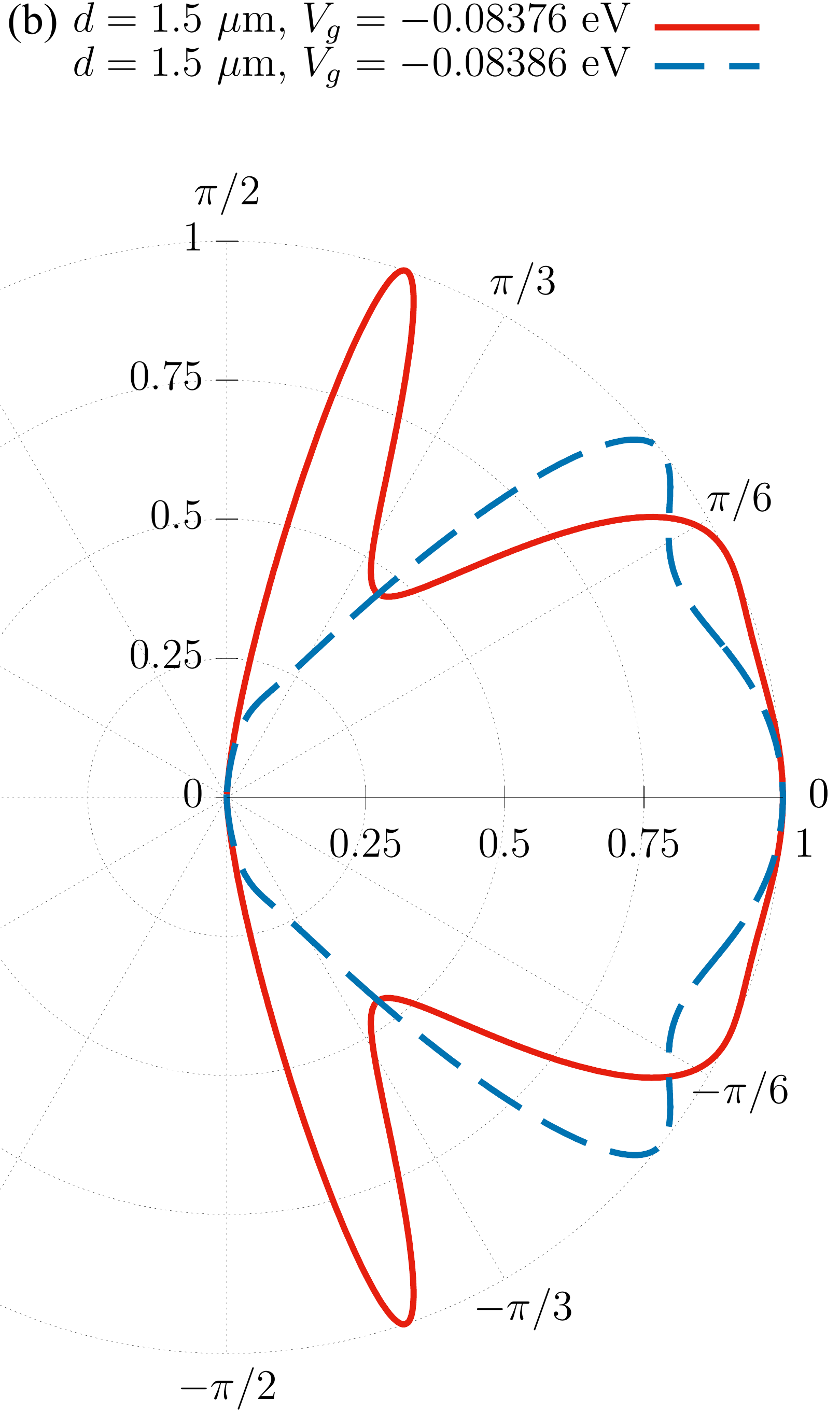}}
\caption{ {Polar plot of transmission probability as a function of the angle of incidence for different width of the gated region, $d$. $V_g$ are chosen in the oscillatory regime and red and blue curve correspond to the closest maxima and minima of the conductance.  We used $\mu=0.02$ eV, $v_1 = 3.3$ eV {\AA}, $v_2 = 1.65$ eV {\AA}, $\alpha = 0$ and $\beta = 0$.}}
\label{fig:polar}
\end{figure}

The general behavior of the transmission coefficients for different choices of $k_Fd$ are shown in Fig.~\ref{fig:polar}. Note that for smaller gated length $T(\theta)$ is generally featureless, with maximum at normal incidence. As the gate voltage is swept from minimum to the maximum of the conductance, the transmission coefficient at every angle changes accordingly, see Fig.~\ref{fig:polar}(a), in agreement with Eq.~\eqref{trans_simple}. For larger $k_Fd$, Fig.~\ref{fig:polar}(b), transmission maxima appear at angles $\theta_n$. These angles are shifted as the gate voltage varies, leading to small amplitude of conductance oscillations in this regime.

\begin{figure}[t]
\centering
\subfloat{\includegraphics[width=0.8\columnwidth]{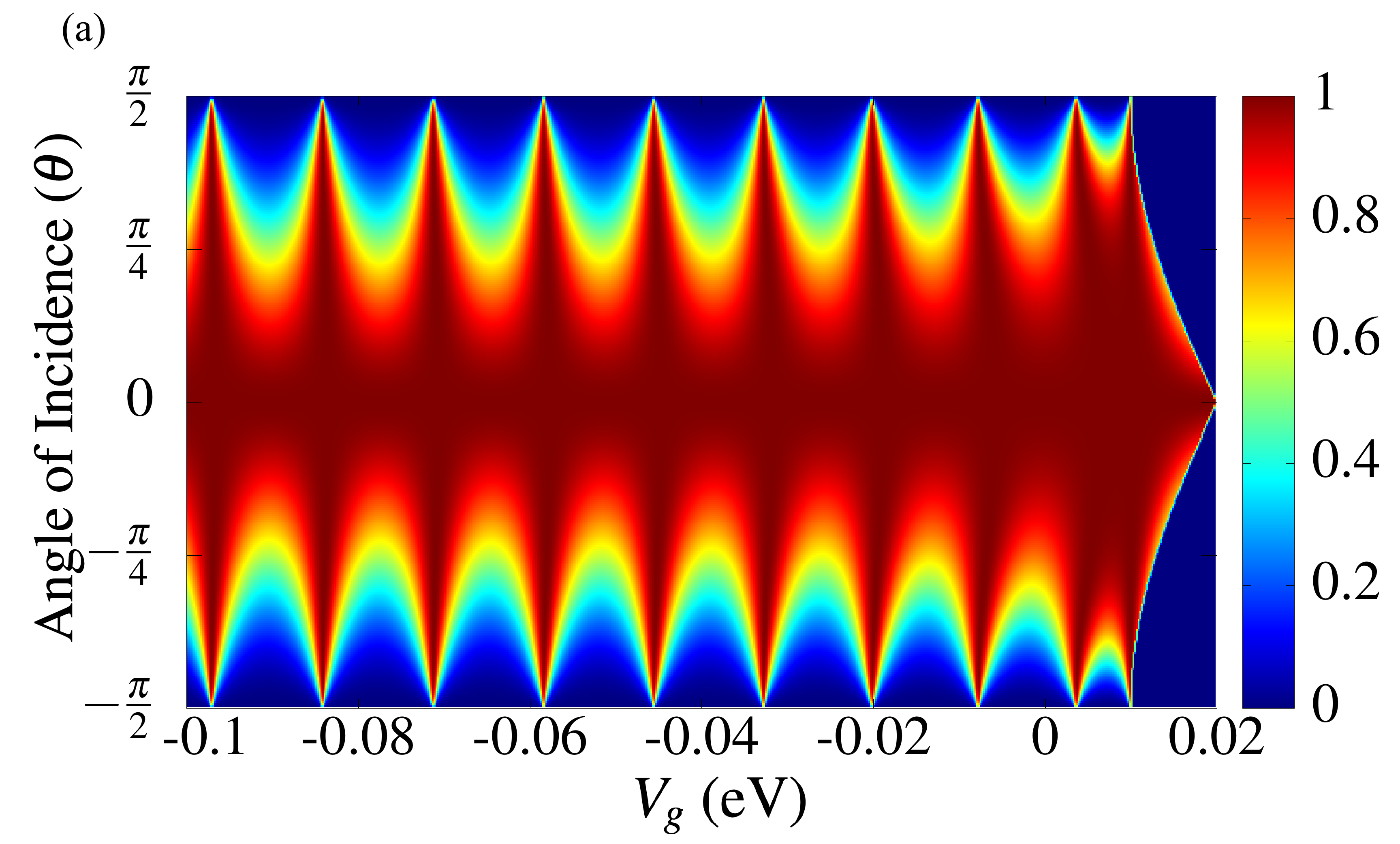}}
\hspace{0.2cm}
\subfloat{\includegraphics[width=0.8\columnwidth]{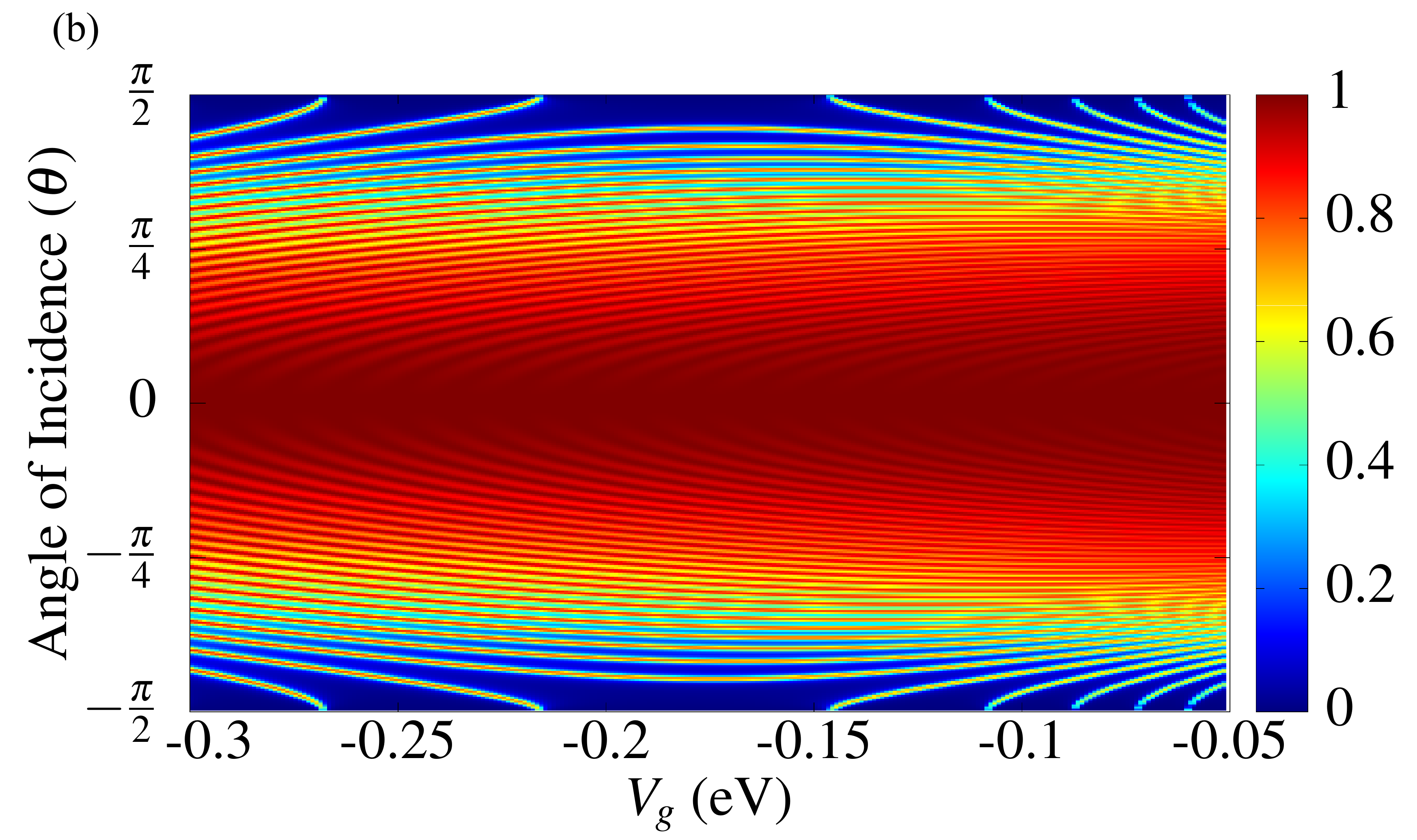}}
\caption{ {Transmission probability as a function of the angle of incidence and the gate voltage $V_g$ for $v_1 = 3.3$ eV {\AA}, $v_2 = 1.65$ eV {\AA} and $\alpha = \beta = 0$ (a) $\mu=0.02$ eV, $d = 40$ nm (b) $\mu = 0.2$ eV, $d = 10$ $\mu$m.}
\label{fig:fringes}}
\end{figure}

The entire behavior of the conductance as a function of both voltage and the incidence angle is summarized in Fig.~\ref{fig:fringes}. At small to moderate values of $k_Fd$, Fig.~\ref{fig:fringes}(a), the periodic pattern or large vs small transmission at finite incidence angles is very clear, with deep troves in between. {As this parameter becomes greater, however, see Fig.~\ref{fig:fringes}(b) where we considered a greater values of $\mu$ and $V_g$ to bring out the features more clearly, multiple transmission peaks are present, the number of such peaks depends on the gate voltage, and the oscillation amplitude, which is the weighted integral of $T(\theta)$, is reduced.}


In summary, here we showed that, given knowledge of the surface state dispersion in a topological insulator, $v_1$, we can determine the Dirac velocity of the quasiparticles in the (inaccessible to surface probes) interface region from the behavior of the conductance of a double junction using either the kink voltage, Eq.~\eqref{kink_V}, or period of the oscillations, Eq.~\eqref{V_g_period}. For the current generation of topological insulators such as Bi$_2$Se$_3$, to stay well below the bulk gap we estimate the maximal value of the gate voltage, $V_g\lesssim 0.13$ eV (taking half of the gap value as the rough guide), implying that the chemical potential should stay below about half of that value to ensure sufficient range of oscillations,  $\mu\lesssim 0.06$eV, translating into $k_F \leq 2 \times 10^{-2}$\AA. Therefore, the values of roughly $d\lesssim 10^3$\AA \ correspond to small gate, while $d\geq 5000$\AA \ correspond to large gate regimes. The nature and origin of the oscillations are slightly distinct between the two regimes, and we provided both analytical and numerical evaluations of both.

Crucially, our results are insensitive to the boundary scattering at lateral junctions, and therefore applicable for a wide range of techniques of sample preparation. We now show that this seeming redundancy in determining $v_2$ can be used to quantitatively determine the anisotropy of the interface Dirac states when the symmetry is broken in the interface region.

\section{Non-helical interface states}
\label{sec:spinz}

The above analysis was for a system with the simplest possible effect of the interface: velocity renormalization of the helical states. The key question is whether the interface states with broken rotational symmetry (and therefore non-helical) can be detected and characterized by the proposed method. The answer is affirmative, and we give the details here.
Ref.~\onlinecite{mahmoud_se_ti} gave the symmetry analysis of all possible interface state Hamiltonians linear in $\bm k$, see Eq.~\eqref{Ham0}, and we focus on the most interesting case considered there. When the spin-rotation invariance is broken, the residual symmetry of $H_I$ is reduced to the ${\cal B}_2$ representation of the $C_{2v}$ symmetry group, the Dirac dispersion becomes anisotropic, with elliptical constant energy contours, and the spins of the interface states point out of the plane of the interface. To capture these effects, we consider
\begin{equation}
  H_I = v_2(\sigma_x k_y - a\sigma_y k_x)+ v_2 b\sigma_z k_x\,,
  \label{HI_eff}
\end{equation}
equivalent to that analyzed in Ref.~\onlinecite{mahmoud_se_ti}. In simple models
the values of $a,b$ depend on the interface potentials, and generically, without fine tuning, $b\neq 0$ whenever $a\neq 1$. For $a=1,b=0$ we recover the helical interface state case discussed in the previous section). The energy dispersion of $H_I$,
\begin{equation}
E_\pm(k_x,k_y)=\pm v_2\sqrt{(a^2+b^2)k_x^2+ k_y^2},
\label{energy_non_hel}
\end{equation}
is anisotropic, with elliptical constant energy contours.

It is convenient to introduce the following notations,
\begin{subequations}
  \begin{eqnarray}
  &&k=\sqrt{(a^2+b^2)k_x^2+k_y^2}\,, 
  \\
  &&\tan\eta=\frac{k_y}{\sqrt{a^2+b^2}k_x}\,,
  \\
  &&\tan\psi=\frac{b}{a}\,.
\end{eqnarray}
 \label{wavefn_params}
 \end{subequations}
Then the eigenvector of the interface hamiltonian, $H_I+V_g\openone$, corresponding to the eigenvalue, $E_+ + V_g$, is given by
\begin{eqnarray}
  \widetilde\Psi_I(E_++V_g,\bm k)&=&\frac{1}{\sqrt{2-2\cos\eta\sin\psi}}
  \\
  \nonumber
  &&\times
  \begin{pmatrix}
    \sin\eta+i\cos\eta\cos\psi\\
    1-\cos\eta\sin\psi\\
  \end{pmatrix}\,.
\end{eqnarray}



The eigenstate is non-helical, and the spins tilt out of the plane of the interface, with the maximal out-of-plane polarization is $b/\sqrt{a^2+b^2}=\sin\psi$ at $k_y=0$.
In Eq.~\eqref{HI_eff} we made the natural assumption that symmetry breaking (e.g. due to strain) yields the principal axes for the ellipse which are parallel and perpendicular to the interface. The spins acquire a non-zero out-of-plane component for $b\neq 0$.

As discussed in the previous sections, for each choice of the interface Hamiltonian, we need to derive the appropriate boundary conditions, conserving the particle current normal to the interface. Here at the left interface in Fig.~\ref{fig:device}, we have $j_x^-=-v_1\sigma_y$, while $j_x^+ = v_2(-a\sigma_y+b\sigma_z)$, so that the matrix $M_L$ satisfying Eq.~\eqref{M_cur} has the form
\begin{eqnarray}
    M_L(\alpha)&=&\sqrt{\frac{v_2}{v_1}}\left[a^2+b^2\right]^{1/4}
    \\
    \nonumber
    &&\times\left[\cos\frac{\psi}{2}e^{i\alpha\sigma_y}-i\sin\frac{\psi}{2}\sigma_x e^{-i\alpha\sigma_y}\right]\,.
	\label{M_B}
\end{eqnarray}
Here $\alpha$ is the parameter encoding the potential scattering at the junction. Similarly, for the right junction $M_R(\beta) = [M_L(-\beta)]^{-1}$. We then compute the transmission coefficient and the conductance as described above.

\begin{figure}[t]
\includegraphics[width=\columnwidth]{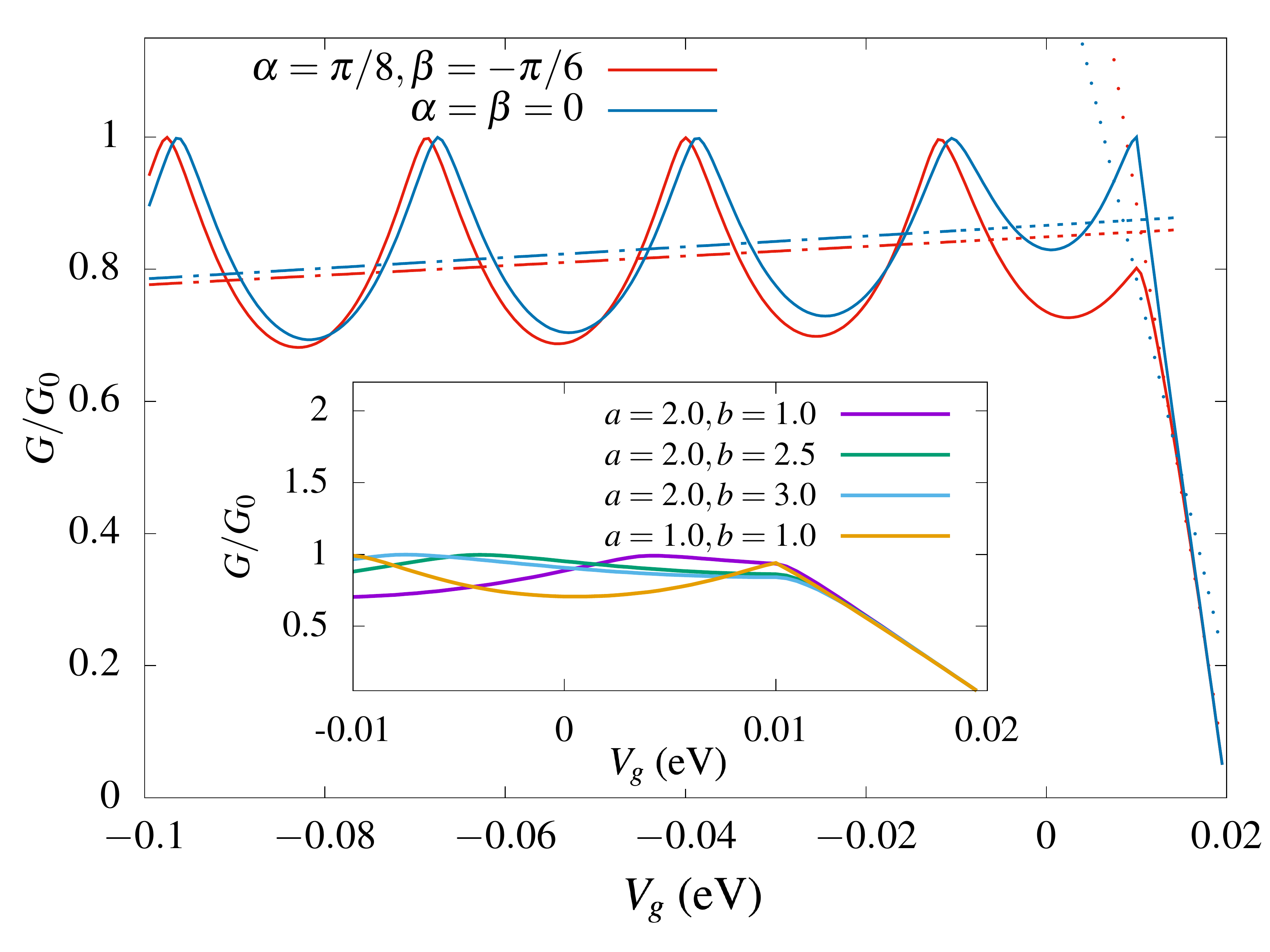}
\caption{{Conductance for the non-helical interface state for $v_2/v_1=1/2$. Main panel: Determination of the kink voltage. Dot-dashed lines: average of oscillatory $G(V_g)$; dashed line: extrapolation; arrows indicate $V_k^{(-)}$. Dotted line: slope of $G(V_g)$. Inset: Independence of the initial slope and the kink voltage of the conductance on the ellipticity and non-helicity parameters $a,b$.}}
\label{fig:G_B}
\end{figure}

The main features of the conductance as a function of $V_g$, shown in Fig.~\ref{fig:G_B}, are similar to those for the simple interface considered in previous section. When the Dirac point is close to the chemical potential $G(V_g)$ is quasi-linear due to the reduction in the density of states. For large mismatch of the Fermi surfaces between the surface and the interface regions, we observe the conductance oscillations. However, the two regimes are now controlled by the Fermi velocities in different directions.

Recall that the kink voltage, $V_k^{(\pm)}$, corresponds to the maximal magnitude of $|V_g-\mu|$ for which the conservation of the momentum along the junction still allows transmission for any incidence angle, see Eq.~\eqref{k_y_cons}. As the quasiparticle velocity along the junction in the interface region is $v_2$, see Eq.~\eqref{energy_non_hel}, the kink voltage depends only on that value, and is still given by Eq.~\eqref{kink_V}.
Fig.~\ref{fig:G_B} indeed shows that the kink voltage obtained from the full numerical evaluation is in agreement with that value, and only weakly depends on the junction scattering parameters, $\alpha,\beta$. In contrast, the initial slope, $dG/dV_g\|_{V_g=\mu}$, is more sensitive to these parameters. Fig.~\ref{fig:G_B}(b) also confirms that the kink voltage is completely insensitive to the choice of the anisotropy parameters $a$ and $b$.

\begin{figure}[t]
\includegraphics[width=0.5\textwidth]{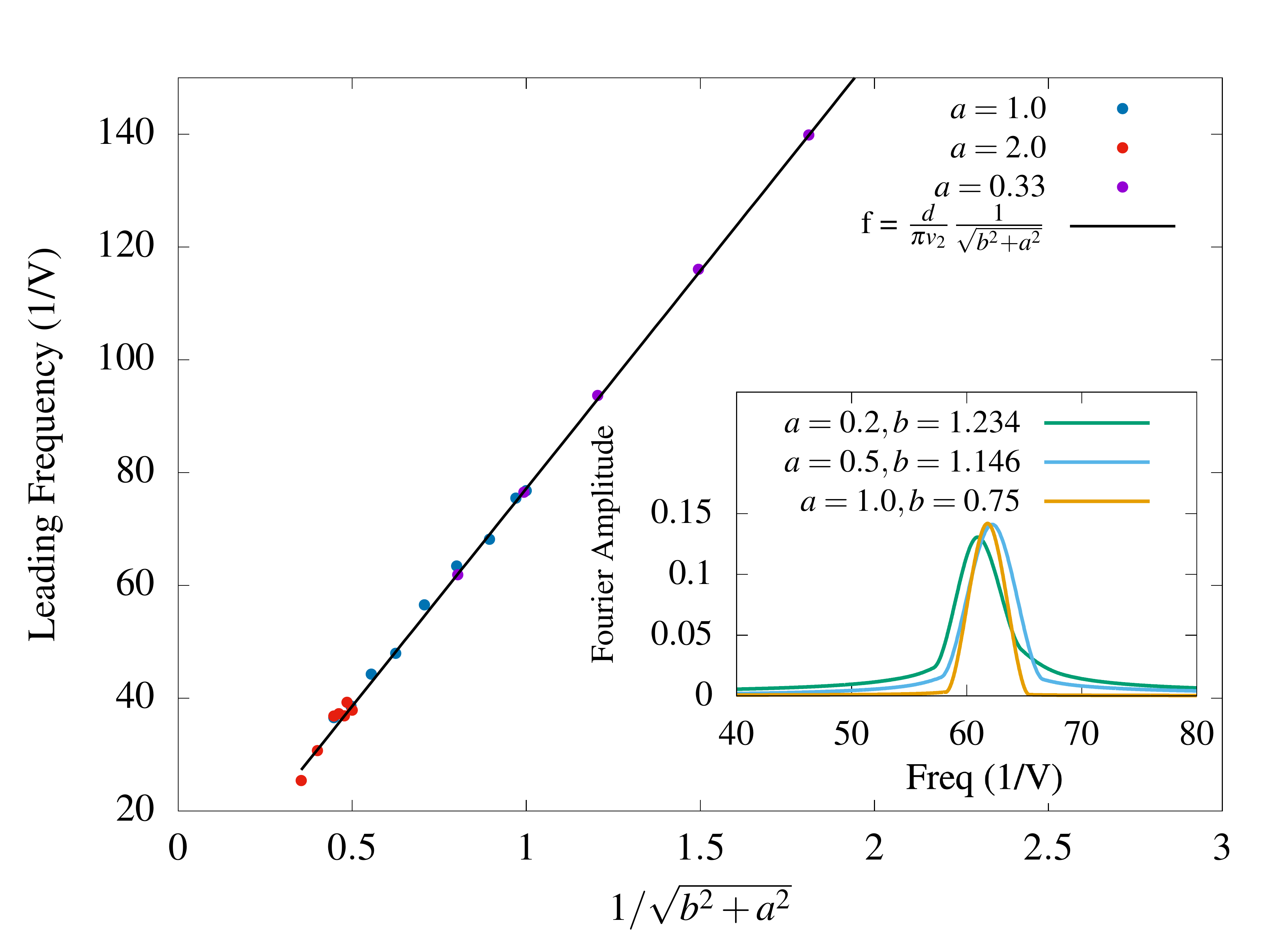}
\caption{{Leading frequency obtained by Fourier transform of the conductance oscillations depends only on $v_{F,x}^{max}$, for various values of $a,b$. The relation of this frequency to $a,b$, and thus to $v_{F,x}^{max}$, is shown in Eq.~\eqref{freq_nonhel}. Inset: Same lead frequency for different values of $a,b$.}}
\label{fig:lead_freq}
\end{figure}

In the oscillatory regime, as discussed in the previous section, a good approximation for the conductance oscillations is obtained by assuming that in the interface region the quasiparticles propagate nearly normal to the junction. Therefore the relevant velocity is $v_{F,x}^{max}=v_2\sqrt{a^2+b^2}$, and we expect the leading oscillation frequency to be given by the corresponding modification of Eq.~\eqref{V_g_period}, namely,
\begin{equation}
  f = \frac{d}{\pi v_{F,x}^{max}} = \frac{d}{\pi v_2\sqrt{a^2+b^2}} \,.
  \label{freq_nonhel}
\end{equation}
Indeed, the lead frequency of the Fourier transform of the conductance oscillations, plotted in Fig.~\ref{fig:lead_freq} shows an excellent agreement with this expression. It is also essentially independent of the junction scattering.

Consequently, in the more general case considered here, measurements of the kink voltage and the lead frequency of the conductance oscillations give the values for the quasiparticle velocities in the interface region along and normal to the interface respectively, this giving a quantitative measure of the anisotropy (or lack thereof) of the topological states in the otherwise inaccessible interface region. To our knowledge this is the first proposal for such a measurement.

\section{Discussion and conclusions}
\label{sec:disc}

In summary, we addressed here a problem of general interest for prototypical topological devices. The functionality of such devices relies on interface, rather than the surface, topological states, but the interface potentials may break the full rotational symmetry of the low-energy topological states already at the level of linear, in the momentum, terms in the effective Hamiltonian. Given that the interface states are not easily accessible by spectroscopic probes such as STM or ARPES, the question of how to detect this broken symmetry is relevant and timely. Symmetry breaking of the interface states has implications for a wide variety of devices of interest that depend on using topological states from  TI based ferromagnetic devices to TI based Josephson junctions which have been shown to host the famous zero-energy Majorana fermions.

We showed that the conductance of a gated mesoscale double junction at a surface of topological insulator can be used to quantitatively determine the anisotropy (or lack thereof) of the quasiparticle dispersion normal to and along the junction in the (otherwise hidden) interface region. We demonstrated that the features we use to extract these parameters, the kink and the period of Fabri-Perot-like oscillations of conductance as a function of gate voltage, are robust against scattering at the junctions, and independent of the exact location of the chemical potential, provided the bulk band contribution to the conductance and other experimentally controlled extrinsic effects remain small.

While the broken rotational symmetry of the interface states indicates that these states are non-helical, and, generally, in the absence of fine-tuning, requires out of plane spin-tilt of the topological states, the measurement we propose does not directly detect this out of plan spin component. A different set of measurements, possibly under applied magnetic field, is required to do this, and will be addressed separately. However, observation of the anisotropy in the quasiparticle dispersion for the topological states would in itself be an important step towards understanding, further analyzing from the first-principles perspective, and ultimately controlling, the topological interface properties for prototype devices.

\begin{acknowledgments}
We thank D. Alspaugh and D. E. Sheehy for discussions. This research was supported in part by NSF via Grant No. DMR-1410741 (E.~T., I.~V., M.~M.~A.) and No. DMR-1151717 (M.~M.~A.).
\end{acknowledgments}

\appendix

\section{Helical Interface State}
\label{sec:app_hel}

In this appendix, we detail the $\mathcal{T}$ and $\mathcal{S}$-matrix based method used to calculate transmission probability.

To begin we recall that the Hamiltonian of the gated double junction is given by

\begin{equation}
H =
\begin{cases}
v_1({\bm \sigma} \times {\bm k}), & \text{if } x \leq 0 \text{ Region I }\\
v_2({\bm\sigma} \times {\bm k'})+V_g, & \text{if } 0< x < d \text{ Region II }\\
v_1({\bm\sigma} \times {\bm k}), & \text{if } x\geq d \text{ Region III }
\end{cases}
\end{equation}
We can write the eigenvectors for positive energy $E$, in different regions  as
\begin{equation}\label{RI}
\Psi_I = \frac{A}{\sqrt{2}} e^{i(k_xx+k_yy)}
\begin{pmatrix}
i \\
e^{i\theta}
\end{pmatrix}
+\frac{B}{\sqrt{2}}e^{i(-k_xx+k_yy)}
\begin{pmatrix}
i \\
-e^{-i\theta}
\end{pmatrix},
\end{equation}
\begin{equation}
\Psi_{II} = \frac{C}{\sqrt{2}} e^{i(k_x'x+k_y'y)}
\begin{pmatrix}
i \\
se^{i\phi}
\end{pmatrix}
+\frac{D}{\sqrt{2}}e^{i(-k_x'x+k_y'y)}
\begin{pmatrix}
i \\
-se^{-i\phi}
\end{pmatrix},
\end{equation}

\begin{equation}\label{RIII}
\Psi_{III} = \frac{E}{\sqrt{2}} e^{i(k_xx+k_yy)}
\begin{pmatrix}
i \\
e^{i\theta}
\end{pmatrix}
+\frac{F}{\sqrt{2}}e^{i(-k_xx+k_yy)}
\begin{pmatrix}
i \\
-e^{-i\theta}
\end{pmatrix}.
\end{equation}
In the latter equations we used $k_y' = k_y$ due to translational invariance along the $y$-direction and defined $s =$ sgn ${(\mu-V_g)}$. Before preceding to solve the boundary conditions at $x=0$ and $x=d$ junction it is convenient to define the matrix,
\begin{equation}\label{matrixp}
P(k,\theta,x,s) =\frac{1}{\sqrt{2}}
\begin{pmatrix}
ie^{ik\cos\theta x} & ie^{-ik\cos\theta x}\\
se^{i\theta}e^{ik\cos\theta x} & -se^{-i\theta}e^{-ik\cos\theta x}
\end{pmatrix}\;.
\end{equation}
We notice that for different parameters $x$, $\theta$ and $s$ in the matrix $P$ make its columns the eigenvector components in the different scattering regions. Returning to the boundary condition at the first junction, $x=0$, we notice that
\begin{multline}
M_L(\alpha)\Psi(0_+,y) = \Psi(0_-,y)\\
M_L^{-1}(\alpha)P(k,\theta,0,1)
\begin{pmatrix}
A\\
B
\end{pmatrix}
=
P(k',\phi,0,s)
\begin{pmatrix}
C\\
D
\end{pmatrix}\;.
\end{multline}
Similarly, at the second junction ($x=d$),
\begin{multline}
M_R(\beta)\Psi_{III}(d_+,y) = \Psi_{II}(d_-,y)\\
M_R^{-1}(\beta)P(k',\phi,d,s)
\begin{pmatrix}
C\\
D
\end{pmatrix}
=
P(k,\theta,d,1)
\begin{pmatrix}
E\\
F
\end{pmatrix}\;.
\end{multline}

Using the above two boundary conditions one can write
\begin{multline}
\begin{pmatrix}
E\\
F
\end{pmatrix}=
P^{-1}(k,\theta,d,1) M_R^{-1}(\beta) P(k',\phi,d,s)\\P^{-1}(k',\phi,0,s) M_L^{-1}(\alpha) P(k,\theta,0,1)\begin{pmatrix}
A\\
B
\end{pmatrix}\;,
\end{multline}
and from the equation above we can identify the transmission-matrix ($\mathcal{T}$-matrix), where
\begin{eqnarray}
\mathcal{T}=&&P^{-1}(k,\theta,d,1) M_R^{-1}(\beta) P(k',\phi,d,s)\nonumber\\ &&\times P^{-1}(k',\phi,0,s) M_L^{-1}(\alpha) P(k,\theta,0,1)\;.
\end{eqnarray}
Here we note that the presence of the junction-potentials modify the $\mathcal{T}$-matrix, from its conventional form with $M_L^{-1}(\alpha)=M_R^{-1}(\beta)=I$, where $I$ is the identity matrix), via the matrices $M_L^{-1}(\alpha)$ and $M_R^{-1}(\beta) $ which encapsulate the scattering sources via $\alpha$ and $\beta$ at their respective junctions. The $\mathcal{T}$-matrix is related to the scattering-matrix ($\mathcal{S}$-matrix) by
\begin{equation}
\begin{pmatrix}
\mathcal{S}_{11} & \mathcal{S}_{12}\\
\mathcal{S}_{21} &\mathcal{S}_{22}
\end{pmatrix}
=
\begin{pmatrix}
-\mathcal{T}_{21}/\mathcal{T}_{22} & \mathcal{T}_{12}/T_{22}\\
1/\mathcal{T}_{11}^* & \mathcal{T}_{12}/\mathcal{T}_{22}
\end{pmatrix}\;.
\end{equation}
The elements of the $\mathcal{S}$-matrix allow us to determine the transmission and reflection coefficients of the multi-junction scattering problem, such that $r = \mathcal{S}_{11}$ and $t = \mathcal{S}_{21}$. Finally, the transmission probability is given by $T=1-|r|^2=1-|\mathcal{S}_{11}|^2$.

\section{Non-Helical Interface State}

In this section we detail the method used to calculate the transmission probability
for the junctions with non-helical state. The Hamiltonian for our setup is given by

\begin{multline}
H =
\begin{cases}
v_1({\bm \sigma} \times {\bm k}), & \text{if } x \leq 0 \text{ Region I }\\
v_2(\sigma_x k'_y - a\sigma_y k'_x) & \\
+ v_2 b\sigma_z k'_x +V_g,         &\text{if } 0< x < d \text{ Region II }\\
v_1({\bm \sigma} \times {\bm k}), & \text{if } x\geq d \text{ Region III.}
\end{cases}
\end{multline}

In the main text, eigenenergy of region II was written as $\tilde{E}_\pm(k_x',k_y')=\pm v_2\sqrt{(a^2+b^2)k_x'^2+ k_y'^2} +V_g = E_{\pm}(k_x',k_y') +V_g$ and the energy eigenfunction as
\begin{equation}
  \widetilde{\Psi}_{II}(E_\pm,\bm k')=\frac{-i\sin\eta+\cos\eta\cos\psi}{\sqrt{\epsilon^2-2\epsilon\cos\eta\sin\psi +1}}
  \begin{pmatrix}
    i\\
    \frac{\epsilon-\cos\eta\sin\psi}{-i\sin\eta+\cos\eta\cos\psi}\\
  \end{pmatrix}\,.
\end{equation}

where $\epsilon=\frac{E_\pm(k_x,k_y)}{v_2k}= \pm 1$ and $\eta$, $\psi$ are defined in Eq.~\eqref{wavefn_params}.
While this notation is analytically convenient, we switched to a different notation while calculating transmission probability for numerical efficiency. We outline the steps below. Consider the expressions below rewritten using Eq.~\eqref{wavefn_params}

\begin{equation}
\frac{\epsilon-\cos\eta\sin\psi}{-i\sin\eta+\cos\eta\cos\psi} = \frac{k'\epsilon-bk_x'}{ak_x'-ik_y'}
\label{exp1_simp1}
\end{equation}

\begin{equation}
\frac{-i\sin\eta+\cos\eta\cos\psi}{\sqrt{\epsilon^2-2\epsilon\cos\eta\sin\psi +1}} = \frac{ak_x'-ik_y'}{\sqrt{k'}\sqrt{2k'-2\epsilon b k_x'}}
\label{exp2_simp1}
\end{equation}

We now define $K = \sqrt{(a k'_x)^2+k_y'^2}$, $\varphi$ by $e^{i\varphi} = (ak'_x + i k'_y)/K$, so $\tilde{E}_{\pm} = \pm v_2 K \sqrt{[b\cos^2(\varphi)/a]^2 +1}+ V_g$. Using these we can rewrite Eq.~\eqref{exp1_simp1} and \eqref{exp2_simp1} as

\begin{equation}
\frac{\epsilon-\cos\eta\sin\psi}{-i\sin\eta+\cos\eta\cos\psi} = e^{i\varphi} \Big[\frac{\tilde{E}_\pm-V_g}{v_2 K}-\frac{b}{a}\cos\varphi\Big]
\end{equation}

\begin{equation}
\frac{-i\sin\eta+\cos\eta\cos\psi}{\sqrt{\epsilon^2-2\epsilon\cos\eta\sin\psi +1}} = \frac{e^{i\varphi}}{\sqrt{2}\sqrt{\frac{\tilde{E}_\pm-V_g}{v_2 K}}\sqrt{\frac{\tilde{E}_\pm-V_g}{v_2 K \epsilon}-\frac{b}{a}\epsilon\cos\varphi}}
\end{equation}

With these definitions we can write the non helical Hamiltonian $H_{{\rm II}}$  as:
\begin{equation}
H_{\rm II} = \begin{pmatrix}
v_2 \frac{b}{a} K \cos{\varphi} +V_g & ie^{-i\varphi}v_2 K\\
-ie^{i\varphi} K v_2 & -v_2 \frac{b}{a} K \cos{\varphi} +V_g
\end{pmatrix}
\end{equation}
The latter Hamiltonian is characterized by the eigenvalues $\tilde{E}_{\pm} = \pm v_2 K \sqrt{[b\cos^2(\varphi)/a]^2 +1}+ V_g$. Considering the positive eigenenergies, its associated eigenspinor for $\tilde{E}_{+} = v_2 K \sqrt{[b\cos^2(\varphi)/a]^2 +1}+ V_g$  is given by
\begin{equation}
\widetilde{\Psi}_+(\varphi,b/a)=N_+(\varphi,b/a)\begin{pmatrix}
i\\
e^{i\varphi} \big[\frac{\tilde{E}_{+}-V_{g}}{v_2 K}-\frac{b}{a}\cos\varphi\big]
\end{pmatrix}
\end{equation}
where
\begin{equation} \label{npm}
N_+(\varphi,b/a) = \frac{1}{\sqrt{2}\sqrt{\frac{\tilde{E}_{+}-V_{g}}{v_2 K}}  \Big(\frac{\tilde{E}_{+}-V_{g}}{v_2 K} -\frac{b}{a}\cos\varphi\Big)^{1/2}}
\end{equation}

In the regions I and III the eigenstates are given in Eqs.~\ref{RI} and \ref{RIII}, respectively, while in the region II the eigenstates are
\begin{eqnarray}
\Psi_{II}&&=C\widetilde{\Psi}_{i}e^{ik'(\cos\gamma x + \sin\gamma y)}
            +D\widetilde{\Psi}_{r}e^{ik'(-\cos\gamma x + \sin\gamma y)}\;,\nonumber\\
\end{eqnarray}
where we have defined
\begin{equation}\label{psipms}
\widetilde{\Psi}_{i}= \widetilde{\Psi}_+(\varphi,b/a), \hspace{2mm} \widetilde{\Psi}_{r}= \widetilde{\Psi}_+(\pi-\varphi,b/a)
\end{equation}
$N_{+}$ is given in Eq.~\eqref{npm}, $\gamma$ corresponds to the angle between momentum $\bm {k'}$ and the $x$-axis. The eigenvector with coefficient $C$ is travelling towards right while the one with coefficient $D$ is travelling towards left after reflecting from region III.

In this case it is we need to define an independent matrix for the non helical region with columns consisting of the eigenstates in this region, where
\begin{eqnarray}
\mathcal{P}(x)=\left(\widetilde{\Psi}_{i}e^{ik'x\cos{\gamma}}\;\; \widetilde{\Psi}_{r}e^{-ik'x\cos{\gamma}}\right)\;,
\end{eqnarray}
where $\widetilde{\Psi}_{i,r}$ are given in Eq.~\eqref{psipms}. With this definition the boundary condition at the first junction at $x=0$ becomes
\begin{equation}\label{x0}
M(\alpha)\mathcal{P}(0)\begin{pmatrix}
C\\
D
\end{pmatrix}=P(k,\theta,0,1)\begin{pmatrix}
A\\
B\;.
\end{pmatrix}
\end{equation}
where matrix $P$ is defined in Eq.~\eqref{matrixp}. Similarly, at the second junction $x=d$,
\begin{equation}\label{xd2}
M(\beta)\mathcal{P}(d)\begin{pmatrix}
C\\
D
\end{pmatrix}=P(k,\theta,d,1)\begin{pmatrix}
E\\
F
\end{pmatrix}\;.
\end{equation}

Finally, we can relate the coefficients at the two ends the system, such that
\begin{eqnarray}
\begin{pmatrix}
E\\
F
\end{pmatrix}=&&P^{-1}(k,\theta,d,1)M(\beta)\mathcal{P}(d)\mathcal{P}^{-1}(0)\nonumber \\
            &&\times M^{-1}(\alpha)P (k,\theta,0,1)\begin{pmatrix}
A\\
B
\end{pmatrix}\;,
\end{eqnarray}
and this implies that the $\mathcal{T}$-matrix is
\begin{eqnarray}
  \mathcal{T}=&&P^{-1}(k,\theta,d,1)M(\beta)\mathcal{P}(d)\mathcal{P}^{-1}(0)M^{-1}(\alpha)P(k,\theta,0,1).\nonumber\\
\end{eqnarray}
From the $\mathcal{T}$-matrix one can obtain the $\mathcal{S}$-matrix as previously shown. Then the transmission probability is given by $T = 1-|\mathcal{S}_{22}|^2$. By integrating the transmission probability, one can obtain the conductance of the device, which as shown in the main text.


\begin{thebibliography}{40}%
\makeatletter
\providecommand \@ifxundefined [1]{%
 \@ifx{#1\undefined}
}%
\providecommand \@ifnum [1]{%
 \ifnum #1\expandafter \@firstoftwo
 \else \expandafter \@secondoftwo
 \fi
}%
\providecommand \@ifx [1]{%
 \ifx #1\expandafter \@firstoftwo
 \else \expandafter \@secondoftwo
 \fi
}%
\providecommand \natexlab [1]{#1}%
\providecommand \enquote  [1]{``#1''}%
\providecommand \bibnamefont  [1]{#1}%
\providecommand \bibfnamefont [1]{#1}%
\providecommand \citenamefont [1]{#1}%
\providecommand \href@noop [0]{\@secondoftwo}%
\providecommand \href [0]{\begingroup \@sanitize@url \@href}%
\providecommand \@href[1]{\@@startlink{#1}\@@href}%
\providecommand \@@href[1]{\endgroup#1\@@endlink}%
\providecommand \@sanitize@url [0]{\catcode `\\12\catcode `\$12\catcode
  `\&12\catcode `\#12\catcode `\^12\catcode `\_12\catcode `\%12\relax}%
\providecommand \@@startlink[1]{}%
\providecommand \@@endlink[0]{}%
\providecommand \url  [0]{\begingroup\@sanitize@url \@url }%
\providecommand \@url [1]{\endgroup\@href {#1}{\urlprefix }}%
\providecommand \urlprefix  [0]{URL }%
\providecommand \Eprint [0]{\href }%
\providecommand \doibase [0]{http://dx.doi.org/}%
\providecommand \selectlanguage [0]{\@gobble}%
\providecommand \bibinfo  [0]{\@secondoftwo}%
\providecommand \bibfield  [0]{\@secondoftwo}%
\providecommand \translation [1]{[#1]}%
\providecommand \BibitemOpen [0]{}%
\providecommand \bibitemStop [0]{}%
\providecommand \bibitemNoStop [0]{.\EOS\space}%
\providecommand \EOS [0]{\spacefactor3000\relax}%
\providecommand \BibitemShut  [1]{\csname bibitem#1\endcsname}%
\let\auto@bib@innerbib\@empty
\bibitem [{\citenamefont {Qi}\ and\ \citenamefont {Zhang}(2011)}]{Rev_zhang}%
  \BibitemOpen
  \bibfield  {author} {\bibinfo {author} {\bibfnamefont {X.-L.}\ \bibnamefont
  {Qi}}\ and\ \bibinfo {author} {\bibfnamefont {S.-C.}\ \bibnamefont {Zhang}},\
  }\href {\doibase 10.1103/RevModPhys.83.1057} {\bibfield  {journal} {\bibinfo
  {journal} {Rev. Mod. Phys.}\ }\textbf {\bibinfo {volume} {83}},\ \bibinfo
  {pages} {1057} (\bibinfo {year} {2011})}\BibitemShut {NoStop}%
\bibitem [{\citenamefont {Hasan}\ and\ \citenamefont {Kane}(2010)}]{Rev_kane}%
  \BibitemOpen
  \bibfield  {author} {\bibinfo {author} {\bibfnamefont {M.~Z.}\ \bibnamefont
  {Hasan}}\ and\ \bibinfo {author} {\bibfnamefont {C.~L.}\ \bibnamefont
  {Kane}},\ }\href {\doibase 10.1103/RevModPhys.82.3045} {\bibfield  {journal}
  {\bibinfo  {journal} {Rev. Mod. Phys.}\ }\textbf {\bibinfo {volume} {82}},\
  \bibinfo {pages} {3045} (\bibinfo {year} {2010})}\BibitemShut {NoStop}%
\bibitem [{\citenamefont {Hasan}\ and\ \citenamefont
  {Moore}(2011)}]{Rev_hasan}%
  \BibitemOpen
  \bibfield  {author} {\bibinfo {author} {\bibfnamefont {M.~Z.}\ \bibnamefont
  {Hasan}}\ and\ \bibinfo {author} {\bibfnamefont {J.~E.}\ \bibnamefont
  {Moore}},\ }\href {\doibase 10.1146/annurev-conmatphys-062910-140432}
  {\bibfield  {journal} {\bibinfo  {journal} {Annual Review of Condensed Matter
  Physics}\ }\textbf {\bibinfo {volume} {2}},\ \bibinfo {pages} {55} (\bibinfo
  {year} {2011})}\ \Eprint
  {}
  {} \BibitemShut
  {NoStop}%
\bibitem [{\citenamefont {Mellnik}\ \emph {et~al.}(2014)\citenamefont
  {Mellnik}, \citenamefont {Lee}, \citenamefont {Richardella}, \citenamefont
  {Grab}, \citenamefont {Mintun}, \citenamefont {Fischer}, \citenamefont
  {Vaezi}, \citenamefont {Manchon}, \citenamefont {Kim}, \citenamefont
  {Samarth},\ and\ \citenamefont {Ralph}}]{Mellnik2014}%
  \BibitemOpen
  \bibfield  {author} {\bibinfo {author} {\bibfnamefont {A.~R.}\ \bibnamefont
  {Mellnik}}, \bibinfo {author} {\bibfnamefont {J.~S.}\ \bibnamefont {Lee}},
  \bibinfo {author} {\bibfnamefont {A.}~\bibnamefont {Richardella}}, \bibinfo
  {author} {\bibfnamefont {J.~L.}\ \bibnamefont {Grab}}, \bibinfo {author}
  {\bibfnamefont {P.~J.}\ \bibnamefont {Mintun}}, \bibinfo {author}
  {\bibfnamefont {M.~H.}\ \bibnamefont {Fischer}}, \bibinfo {author}
  {\bibfnamefont {A.}~\bibnamefont {Vaezi}}, \bibinfo {author} {\bibfnamefont
  {A.}~\bibnamefont {Manchon}}, \bibinfo {author} {\bibfnamefont {E.-A.}\
  \bibnamefont {Kim}}, \bibinfo {author} {\bibfnamefont {N.}~\bibnamefont
  {Samarth}}, \ and\ \bibinfo {author} {\bibfnamefont {D.~C.}\ \bibnamefont
  {Ralph}},\ }\href {http://dx.doi.org/10.1038/nature13534} {\bibfield
  {journal} {\bibinfo  {journal} {Nature}\ }\textbf {\bibinfo {volume} {511}},\
  \bibinfo {pages} {449 EP } (\bibinfo {year} {2014})}\BibitemShut {NoStop}%
\bibitem [{\citenamefont {Wang}\ \emph
  {et~al.}(2017{\natexlab{a}})\citenamefont {Wang}, \citenamefont {Zhu},
  \citenamefont {Wu}, \citenamefont {Yang}, \citenamefont {Yu}, \citenamefont
  {Ramaswamy}, \citenamefont {Mishra}, \citenamefont {Shi}, \citenamefont
  {Elyasi}, \citenamefont {Teo}, \citenamefont {Wu},\ and\ \citenamefont
  {Yang}}]{Wang:2017}%
  \BibitemOpen
  \bibfield  {author} {\bibinfo {author} {\bibfnamefont {Y.}~\bibnamefont
  {Wang}}, \bibinfo {author} {\bibfnamefont {D.}~\bibnamefont {Zhu}}, \bibinfo
  {author} {\bibfnamefont {Y.}~\bibnamefont {Wu}}, \bibinfo {author}
  {\bibfnamefont {Y.}~\bibnamefont {Yang}}, \bibinfo {author} {\bibfnamefont
  {J.}~\bibnamefont {Yu}}, \bibinfo {author} {\bibfnamefont {R.}~\bibnamefont
  {Ramaswamy}}, \bibinfo {author} {\bibfnamefont {R.}~\bibnamefont {Mishra}},
  \bibinfo {author} {\bibfnamefont {S.}~\bibnamefont {Shi}}, \bibinfo {author}
  {\bibfnamefont {M.}~\bibnamefont {Elyasi}}, \bibinfo {author} {\bibfnamefont
  {K.-L.}\ \bibnamefont {Teo}}, \bibinfo {author} {\bibfnamefont
  {Y.}~\bibnamefont {Wu}}, \ and\ \bibinfo {author} {\bibfnamefont
  {H.}~\bibnamefont {Yang}},\ }\href {\doibase 10.1038/s41467-017-01583-4}
  {\bibfield  {journal} {\bibinfo  {journal} {Nature Communications}\ }\textbf
  {\bibinfo {volume} {8}},\ \bibinfo {pages} {1364} (\bibinfo {year}
  {2017}{\natexlab{a}})}\BibitemShut {NoStop}%
\bibitem [{\citenamefont {Rakheja}\ \emph {et~al.}(2019)\citenamefont
  {Rakheja}, \citenamefont {Flatt\'e},\ and\ \citenamefont
  {Kent}}]{Rakheja:2019}%
  \BibitemOpen
  \bibfield  {author} {\bibinfo {author} {\bibfnamefont {S.}~\bibnamefont
  {Rakheja}}, \bibinfo {author} {\bibfnamefont {M.~E.}\ \bibnamefont
  {Flatt\'e}}, \ and\ \bibinfo {author} {\bibfnamefont {A.~D.}\ \bibnamefont
  {Kent}},\ }\href {\doibase 10.1103/PhysRevApplied.11.054009} {\bibfield
  {journal} {\bibinfo  {journal} {Phys. Rev. Applied}\ }\textbf {\bibinfo
  {volume} {11}},\ \bibinfo {pages} {054009} (\bibinfo {year}
  {2019})}\BibitemShut {NoStop}%
\bibitem [{\citenamefont {Fu}\ and\ \citenamefont {Kane}(2008)}]{Fu_Kane_2008}%
  \BibitemOpen
  \bibfield  {author} {\bibinfo {author} {\bibfnamefont {L.}~\bibnamefont
  {Fu}}\ and\ \bibinfo {author} {\bibfnamefont {C.~L.}\ \bibnamefont {Kane}},\
  }\href {\doibase 10.1103/PhysRevLett.100.096407} {\bibfield  {journal}
  {\bibinfo  {journal} {Phys. Rev. Lett.}\ }\textbf {\bibinfo {volume} {100}},\
  \bibinfo {pages} {096407} (\bibinfo {year} {2008})}\BibitemShut {NoStop}%
\bibitem [{\citenamefont {Alicea}(2012)}]{Alicea2012}%
  \BibitemOpen
  \bibfield  {author} {\bibinfo {author} {\bibfnamefont {J.}~\bibnamefont
  {Alicea}},\ }\href {http://stacks.iop.org/0034-4885/75/i=7/a=076501}
  {\bibfield  {journal} {\bibinfo  {journal} {Reports on Progress in Physics}\
  }\textbf {\bibinfo {volume} {75}},\ \bibinfo {pages} {076501} (\bibinfo
  {year} {2012})}\BibitemShut {NoStop}%
\bibitem [{\citenamefont {Burkov}\ and\ \citenamefont
  {Hawthorn}(2010)}]{Burkov2010}%
  \BibitemOpen
  \bibfield  {author} {\bibinfo {author} {\bibfnamefont {A.~A.}\ \bibnamefont
  {Burkov}}\ and\ \bibinfo {author} {\bibfnamefont {D.~G.}\ \bibnamefont
  {Hawthorn}},\ }\href {\doibase 10.1103/PhysRevLett.105.066802} {\bibfield
  {journal} {\bibinfo  {journal} {Phys. Rev. Lett.}\ }\textbf {\bibinfo
  {volume} {105}},\ \bibinfo {pages} {066802} (\bibinfo {year}
  {2010})}\BibitemShut {NoStop}%
\bibitem [{\citenamefont {Pesin}\ and\ \citenamefont
  {MacDonald}(2012)}]{Pesin2012}%
  \BibitemOpen
  \bibfield  {author} {\bibinfo {author} {\bibfnamefont {D.}~\bibnamefont
  {Pesin}}\ and\ \bibinfo {author} {\bibfnamefont {A.~H.}\ \bibnamefont
  {MacDonald}},\ }\href {http://dx.doi.org/10.1038/nmat3305} {\bibfield
  {journal} {\bibinfo  {journal} {Nature Materials}\ }\textbf {\bibinfo
  {volume} {11}},\ \bibinfo {pages} {409 EP } (\bibinfo {year}
  {2012})}\BibitemShut {NoStop}%
\bibitem [{\citenamefont {Zhang}\ \emph {et~al.}(2011)\citenamefont {Zhang},
  \citenamefont {Chang}, \citenamefont {Zhang}, \citenamefont {Wen},
  \citenamefont {Feng}, \citenamefont {Li}, \citenamefont {Liu}, \citenamefont
  {He}, \citenamefont {Wang}, \citenamefont {Chen}, \citenamefont {Xue},
  \citenamefont {Ma},\ and\ \citenamefont {Wang}}]{Zhang2011}%
  \BibitemOpen
  \bibfield  {author} {\bibinfo {author} {\bibfnamefont {J.}~\bibnamefont
  {Zhang}}, \bibinfo {author} {\bibfnamefont {C.-Z.}\ \bibnamefont {Chang}},
  \bibinfo {author} {\bibfnamefont {Z.}~\bibnamefont {Zhang}}, \bibinfo
  {author} {\bibfnamefont {J.}~\bibnamefont {Wen}}, \bibinfo {author}
  {\bibfnamefont {X.}~\bibnamefont {Feng}}, \bibinfo {author} {\bibfnamefont
  {K.}~\bibnamefont {Li}}, \bibinfo {author} {\bibfnamefont {M.}~\bibnamefont
  {Liu}}, \bibinfo {author} {\bibfnamefont {K.}~\bibnamefont {He}}, \bibinfo
  {author} {\bibfnamefont {L.}~\bibnamefont {Wang}}, \bibinfo {author}
  {\bibfnamefont {X.}~\bibnamefont {Chen}}, \bibinfo {author} {\bibfnamefont
  {Q.-K.}\ \bibnamefont {Xue}}, \bibinfo {author} {\bibfnamefont
  {X.}~\bibnamefont {Ma}}, \ and\ \bibinfo {author} {\bibfnamefont
  {Y.}~\bibnamefont {Wang}},\ }\href {\doibase 10.1038/ncomms1588} {\bibfield
  {journal} {\bibinfo  {journal} {Nature Communications}\ }\textbf {\bibinfo
  {volume} {2}},\ \bibinfo {pages} {574} (\bibinfo {year} {2011})}\BibitemShut
  {NoStop}%
\bibitem [{\citenamefont {Arakane}\ \emph {et~al.}(2012)\citenamefont
  {Arakane}, \citenamefont {Sato}, \citenamefont {Souma}, \citenamefont
  {Kosaka}, \citenamefont {Nakayama}, \citenamefont {Komatsu}, \citenamefont
  {Takahashi}, \citenamefont {Ren}, \citenamefont {Segawa},\ and\ \citenamefont
  {Ando}}]{Arakane2012}%
  \BibitemOpen
  \bibfield  {author} {\bibinfo {author} {\bibfnamefont {T.}~\bibnamefont
  {Arakane}}, \bibinfo {author} {\bibfnamefont {T.}~\bibnamefont {Sato}},
  \bibinfo {author} {\bibfnamefont {S.}~\bibnamefont {Souma}}, \bibinfo
  {author} {\bibfnamefont {K.}~\bibnamefont {Kosaka}}, \bibinfo {author}
  {\bibfnamefont {K.}~\bibnamefont {Nakayama}}, \bibinfo {author}
  {\bibfnamefont {M.}~\bibnamefont {Komatsu}}, \bibinfo {author} {\bibfnamefont
  {T.}~\bibnamefont {Takahashi}}, \bibinfo {author} {\bibfnamefont
  {Z.}~\bibnamefont {Ren}}, \bibinfo {author} {\bibfnamefont {K.}~\bibnamefont
  {Segawa}}, \ and\ \bibinfo {author} {\bibfnamefont {Y.}~\bibnamefont
  {Ando}},\ }\href {\doibase 10.1038/ncomms1639} {\bibfield  {journal}
  {\bibinfo  {journal} {Nature Communications}\ }\textbf {\bibinfo {volume}
  {3}},\ \bibinfo {pages} {636} (\bibinfo {year} {2012})}\BibitemShut {NoStop}%
\bibitem [{\citenamefont {Zhang}\ \emph {et~al.}(2012)\citenamefont {Zhang},
  \citenamefont {Kane},\ and\ \citenamefont {Mele}}]{FZhang2012}%
  \BibitemOpen
  \bibfield  {author} {\bibinfo {author} {\bibfnamefont {F.}~\bibnamefont
  {Zhang}}, \bibinfo {author} {\bibfnamefont {C.~L.}\ \bibnamefont {Kane}}, \
  and\ \bibinfo {author} {\bibfnamefont {E.~J.}\ \bibnamefont {Mele}},\ }\href
  {\doibase 10.1103/PhysRevB.86.081303} {\bibfield  {journal} {\bibinfo
  {journal} {Phys. Rev. B}\ }\textbf {\bibinfo {volume} {86}},\ \bibinfo
  {pages} {081303(R)} (\bibinfo {year} {2012})}\BibitemShut {NoStop}%
\bibitem [{\citenamefont {Asmar}\ \emph {et~al.}(2017)\citenamefont {Asmar},
  \citenamefont {Sheehy},\ and\ \citenamefont {Vekhter}}]{mahmoud_se_ti}%
  \BibitemOpen
  \bibfield  {author} {\bibinfo {author} {\bibfnamefont {M.~M.}\ \bibnamefont
  {Asmar}}, \bibinfo {author} {\bibfnamefont {D.~E.}\ \bibnamefont {Sheehy}}, \
  and\ \bibinfo {author} {\bibfnamefont {I.}~\bibnamefont {Vekhter}},\ }\href
  {\doibase 10.1103/PhysRevB.95.241115} {\bibfield  {journal} {\bibinfo
  {journal} {Phys. Rev. B}\ }\textbf {\bibinfo {volume} {95}},\ \bibinfo
  {pages} {241115(R)} (\bibinfo {year} {2017})}\BibitemShut {NoStop}%
\bibitem [{\citenamefont {Alspaugh}\ \emph {et~al.}(2018)\citenamefont
  {Alspaugh}, \citenamefont {Asmar}, \citenamefont {Sheehy},\ and\
  \citenamefont {Vekhter}}]{Alspaugh2018}%
  \BibitemOpen
  \bibfield  {author} {\bibinfo {author} {\bibfnamefont {D.~J.}\ \bibnamefont
  {Alspaugh}}, \bibinfo {author} {\bibfnamefont {M.~M.}\ \bibnamefont {Asmar}},
  \bibinfo {author} {\bibfnamefont {D.~E.}\ \bibnamefont {Sheehy}}, \ and\
  \bibinfo {author} {\bibfnamefont {I.}~\bibnamefont {Vekhter}},\ }\href
  {\doibase 10.1103/PhysRevB.98.104516} {\bibfield  {journal} {\bibinfo
  {journal} {Phys. Rev. B}\ }\textbf {\bibinfo {volume} {98}},\ \bibinfo
  {pages} {104516} (\bibinfo {year} {2018})}\BibitemShut {NoStop}%
\bibitem [{\citenamefont {Wang}\ \emph
  {et~al.}(2017{\natexlab{b}})\citenamefont {Wang}, \citenamefont {Zhu},
  \citenamefont {Wu}, \citenamefont {Yang}, \citenamefont {Yu}, \citenamefont
  {Ramaswamy}, \citenamefont {Mishra}, \citenamefont {Shi}, \citenamefont
  {Elyasi}, \citenamefont {Teo}, \citenamefont {Wu},\ and\ \citenamefont
  {Yang}}]{Wang2017}%
  \BibitemOpen
  \bibfield  {author} {\bibinfo {author} {\bibfnamefont {Y.}~\bibnamefont
  {Wang}}, \bibinfo {author} {\bibfnamefont {D.}~\bibnamefont {Zhu}}, \bibinfo
  {author} {\bibfnamefont {Y.}~\bibnamefont {Wu}}, \bibinfo {author}
  {\bibfnamefont {Y.}~\bibnamefont {Yang}}, \bibinfo {author} {\bibfnamefont
  {J.}~\bibnamefont {Yu}}, \bibinfo {author} {\bibfnamefont {R.}~\bibnamefont
  {Ramaswamy}}, \bibinfo {author} {\bibfnamefont {R.}~\bibnamefont {Mishra}},
  \bibinfo {author} {\bibfnamefont {S.}~\bibnamefont {Shi}}, \bibinfo {author}
  {\bibfnamefont {M.}~\bibnamefont {Elyasi}}, \bibinfo {author} {\bibfnamefont
  {K.-L.}\ \bibnamefont {Teo}}, \bibinfo {author} {\bibfnamefont
  {Y.}~\bibnamefont {Wu}}, \ and\ \bibinfo {author} {\bibfnamefont
  {H.}~\bibnamefont {Yang}},\ }\href {\doibase 10.1038/s41467-017-01583-4}
  {\bibfield  {journal} {\bibinfo  {journal} {Nature Communications}\ }\textbf
  {\bibinfo {volume} {8}},\ \bibinfo {pages} {1364} (\bibinfo {year}
  {2017}{\natexlab{b}})}\BibitemShut {NoStop}%
\bibitem [{\citenamefont {Mondal}\ \emph {et~al.}(2018)\citenamefont {Mondal},
  \citenamefont {Saito}, \citenamefont {Aihara}, \citenamefont {Fons},
  \citenamefont {Kolobov}, \citenamefont {Tominaga}, \citenamefont {Murakami},\
  and\ \citenamefont {Hase}}]{Mondal2018}%
  \BibitemOpen
  \bibfield  {author} {\bibinfo {author} {\bibfnamefont {R.}~\bibnamefont
  {Mondal}}, \bibinfo {author} {\bibfnamefont {Y.}~\bibnamefont {Saito}},
  \bibinfo {author} {\bibfnamefont {Y.}~\bibnamefont {Aihara}}, \bibinfo
  {author} {\bibfnamefont {P.}~\bibnamefont {Fons}}, \bibinfo {author}
  {\bibfnamefont {A.~V.}\ \bibnamefont {Kolobov}}, \bibinfo {author}
  {\bibfnamefont {J.}~\bibnamefont {Tominaga}}, \bibinfo {author}
  {\bibfnamefont {S.}~\bibnamefont {Murakami}}, \ and\ \bibinfo {author}
  {\bibfnamefont {M.}~\bibnamefont {Hase}},\ }\href {\doibase
  10.1038/s41598-018-22196-x} {\bibfield  {journal} {\bibinfo  {journal}
  {Scientific Reports}\ }\textbf {\bibinfo {volume} {8}},\ \bibinfo {pages}
  {3908} (\bibinfo {year} {2018})}\BibitemShut {NoStop}%
\bibitem [{\citenamefont {Beenakker}(2008)}]{Beenakker_Andreev_graphene_2008}%
  \BibitemOpen
  \bibfield  {author} {\bibinfo {author} {\bibfnamefont {C.~W.~J.}\
  \bibnamefont {Beenakker}},\ }\href {\doibase 10.1103/RevModPhys.80.1337}
  {\bibfield  {journal} {\bibinfo  {journal} {Rev. Mod. Phys.}\ }\textbf
  {\bibinfo {volume} {80}},\ \bibinfo {pages} {1337} (\bibinfo {year}
  {2008})}\BibitemShut {NoStop}%
\bibitem [{\citenamefont {Castro~Neto}\ \emph {et~al.}(2009)\citenamefont
  {Castro~Neto}, \citenamefont {Guinea}, \citenamefont {Peres}, \citenamefont
  {Novoselov},\ and\ \citenamefont {Geim}}]{CastroNeto_graphene_review_2009}%
  \BibitemOpen
  \bibfield  {author} {\bibinfo {author} {\bibfnamefont {A.~H.}\ \bibnamefont
  {Castro~Neto}}, \bibinfo {author} {\bibfnamefont {F.}~\bibnamefont {Guinea}},
  \bibinfo {author} {\bibfnamefont {N.~M.~R.}\ \bibnamefont {Peres}}, \bibinfo
  {author} {\bibfnamefont {K.~S.}\ \bibnamefont {Novoselov}}, \ and\ \bibinfo
  {author} {\bibfnamefont {A.~K.}\ \bibnamefont {Geim}},\ }\href {\doibase
  10.1103/RevModPhys.81.109} {\bibfield  {journal} {\bibinfo  {journal} {Rev.
  Mod. Phys.}\ }\textbf {\bibinfo {volume} {81}},\ \bibinfo {pages} {109}
  (\bibinfo {year} {2009})}\BibitemShut {NoStop}%
\bibitem [{\citenamefont {Mondal}\ \emph {et~al.}(2010)\citenamefont {Mondal},
  \citenamefont {Sen}, \citenamefont {Sengupta},\ and\ \citenamefont
  {Shankar}}]{Mondal2010}%
  \BibitemOpen
  \bibfield  {author} {\bibinfo {author} {\bibfnamefont {S.}~\bibnamefont
  {Mondal}}, \bibinfo {author} {\bibfnamefont {D.}~\bibnamefont {Sen}},
  \bibinfo {author} {\bibfnamefont {K.}~\bibnamefont {Sengupta}}, \ and\
  \bibinfo {author} {\bibfnamefont {R.}~\bibnamefont {Shankar}},\ }\href
  {\doibase 10.1103/PhysRevLett.104.046403} {\bibfield  {journal} {\bibinfo
  {journal} {Phys. Rev. Lett.}\ }\textbf {\bibinfo {volume} {104}},\ \bibinfo
  {pages} {046403} (\bibinfo {year} {2010})}\BibitemShut {NoStop}%
\bibitem [{\citenamefont {Soori}\ \emph {et~al.}(2012)\citenamefont {Soori},
  \citenamefont {Das},\ and\ \citenamefont {Rao}}]{Soori2012}%
  \BibitemOpen
  \bibfield  {author} {\bibinfo {author} {\bibfnamefont {A.}~\bibnamefont
  {Soori}}, \bibinfo {author} {\bibfnamefont {S.}~\bibnamefont {Das}}, \ and\
  \bibinfo {author} {\bibfnamefont {S.}~\bibnamefont {Rao}},\ }\href {\doibase
  10.1103/PhysRevB.86.125312} {\bibfield  {journal} {\bibinfo  {journal} {Phys.
  Rev. B}\ }\textbf {\bibinfo {volume} {86}},\ \bibinfo {pages} {125312}
  (\bibinfo {year} {2012})}\BibitemShut {NoStop}%
\bibitem [{not()}]{note}%
  \BibitemOpen
  \href@noop {} {\bibinfo {title} {The effect of the in-plane field
  in our model is much more complex, anisotropically gapping the tis spectrum
  and modifying the boundary conditions, and hence deserves a separate
  study}}\ \BibitemShut {NoStop}%

\bibitem [{\citenamefont {Zhang}\ \emph {et~al.}(2009)\citenamefont {Zhang},
  \citenamefont {Liu}, \citenamefont {Qi}, \citenamefont {Dai}, \citenamefont
  {Fang},\ and\ \citenamefont {Zhang}}]{Zhang2009}%
  \BibitemOpen
  \bibfield  {author} {\bibinfo {author} {\bibfnamefont {H.}~\bibnamefont
  {Zhang}}, \bibinfo {author} {\bibfnamefont {C.-X.}\ \bibnamefont {Liu}},
  \bibinfo {author} {\bibfnamefont {X.-L.}\ \bibnamefont {Qi}}, \bibinfo
  {author} {\bibfnamefont {X.}~\bibnamefont {Dai}}, \bibinfo {author}
  {\bibfnamefont {Z.}~\bibnamefont {Fang}}, \ and\ \bibinfo {author}
  {\bibfnamefont {S.-C.}\ \bibnamefont {Zhang}},\ }\href
  {http://dx.doi.org/10.1038/nphys1270} {\bibfield  {journal} {\bibinfo
  {journal} {Nature Physics}\ }\textbf {\bibinfo {volume} {5}},\ \bibinfo
  {pages} {438 EP } (\bibinfo {year} {2009})},\ \bibinfo {note}
  {article}\BibitemShut {NoStop}%
\bibitem [{\citenamefont {McCann}\ and\ \citenamefont
  {Fal’ko}(2004)}]{falko}%
  \BibitemOpen
  \bibfield  {author} {\bibinfo {author} {\bibfnamefont {E.}~\bibnamefont
  {McCann}}\ and\ \bibinfo {author} {\bibfnamefont {V.~I.}\ \bibnamefont
  {Fal’ko}},\ }\href@noop {} {\bibfield  {journal} {\bibinfo  {journal}
  {Journal of Physics: Condensed Matter}\ }\textbf {\bibinfo {volume} {16}},\
  \bibinfo {pages} {2371} (\bibinfo {year} {2004})}\BibitemShut {NoStop}%
\bibitem [{\citenamefont {Basko}(2009)}]{basko}%
  \BibitemOpen
  \bibfield  {author} {\bibinfo {author} {\bibfnamefont {D.~M.}\ \bibnamefont
  {Basko}},\ }\href@noop {} {\bibfield  {journal} {\bibinfo  {journal} {Phys.
  Rev. B}\ }\textbf {\bibinfo {volume} {79}},\ \bibinfo {pages} {205428}
  (\bibinfo {year} {2009})}\BibitemShut {NoStop}%
\bibitem [{\citenamefont {Akhmerov}\ and\ \citenamefont
  {Beenakker}(2008)}]{akhmerov}%
  \BibitemOpen
  \bibfield  {author} {\bibinfo {author} {\bibfnamefont {A.~R.}\ \bibnamefont
  {Akhmerov}}\ and\ \bibinfo {author} {\bibfnamefont {C.~W.~J.}\ \bibnamefont
  {Beenakker}},\ }\href@noop {} {\bibfield  {journal} {\bibinfo  {journal}
  {Phys. Rev. B}\ }\textbf {\bibinfo {volume} {77}},\ \bibinfo {pages} {085423}
  (\bibinfo {year} {2008})}\BibitemShut {NoStop}%
\bibitem [{\citenamefont {Sen}\ and\ \citenamefont {Deb}(2012)}]{Sen2012}%
  \BibitemOpen
  \bibfield  {author} {\bibinfo {author} {\bibfnamefont {D.}~\bibnamefont
  {Sen}}\ and\ \bibinfo {author} {\bibfnamefont {O.}~\bibnamefont {Deb}},\
  }\href {\doibase 10.1103/PhysRevB.85.245402} {\bibfield  {journal} {\bibinfo
  {journal} {Phys. Rev. B}\ }\textbf {\bibinfo {volume} {85}},\ \bibinfo
  {pages} {245402} (\bibinfo {year} {2012})}\BibitemShut {NoStop}%
\bibitem [{\citenamefont {Isaev}\ \emph {et~al.}(2015)\citenamefont {Isaev},
  \citenamefont {Ortiz},\ and\ \citenamefont {Vekhter}}]{Isaev2015}%
  \BibitemOpen
  \bibfield  {author} {\bibinfo {author} {\bibfnamefont {L.}~\bibnamefont
  {Isaev}}, \bibinfo {author} {\bibfnamefont {G.}~\bibnamefont {Ortiz}}, \ and\
  \bibinfo {author} {\bibfnamefont {I.}~\bibnamefont {Vekhter}},\ }\href
  {\doibase 10.1103/PhysRevB.92.205423} {\bibfield  {journal} {\bibinfo
  {journal} {Phys. Rev. B}\ }\textbf {\bibinfo {volume} {92}},\ \bibinfo
  {pages} {205423} (\bibinfo {year} {2015})}\BibitemShut {NoStop}%
\bibitem [{\citenamefont {Tanhayi~Ahari}\ \emph {et~al.}(2016)\citenamefont
  {Tanhayi~Ahari}, \citenamefont {Ortiz},\ and\ \citenamefont
  {Seradjeh}}]{Tanhayi2016}%
  \BibitemOpen
  \bibfield  {author} {\bibinfo {author} {\bibfnamefont {M.}~\bibnamefont
  {Tanhayi~Ahari}}, \bibinfo {author} {\bibfnamefont {G.}~\bibnamefont
  {Ortiz}}, \ and\ \bibinfo {author} {\bibfnamefont {B.}~\bibnamefont
  {Seradjeh}},\ }\href {\doibase 10.1119/1.4961500} {\bibfield  {journal}
  {\bibinfo  {journal} {American Journal of Physics}\ }\textbf {\bibinfo
  {volume} {84}},\ \bibinfo {pages} {858} (\bibinfo {year} {2016})}\ \Eprint
  {}
  {} \BibitemShut {NoStop}%
\bibitem [{\citenamefont {Kuroda}\ \emph {et~al.}(2010)\citenamefont {Kuroda},
  \citenamefont {Arita}, \citenamefont {Miyamoto}, \citenamefont {Ye},
  \citenamefont {Jiang}, \citenamefont {Kimura}, \citenamefont {Krasovskii},
  \citenamefont {Chulkov}, \citenamefont {Iwasawa}, \citenamefont {Okuda},
  \citenamefont {Shimada}, \citenamefont {Ueda}, \citenamefont {Namatame},\
  and\ \citenamefont {Taniguchi}}]{Kuroda2010}%
  \BibitemOpen
  \bibfield  {author} {\bibinfo {author} {\bibfnamefont {K.}~\bibnamefont
  {Kuroda}}, \bibinfo {author} {\bibfnamefont {M.}~\bibnamefont {Arita}},
  \bibinfo {author} {\bibfnamefont {K.}~\bibnamefont {Miyamoto}}, \bibinfo
  {author} {\bibfnamefont {M.}~\bibnamefont {Ye}}, \bibinfo {author}
  {\bibfnamefont {J.}~\bibnamefont {Jiang}}, \bibinfo {author} {\bibfnamefont
  {A.}~\bibnamefont {Kimura}}, \bibinfo {author} {\bibfnamefont {E.~E.}\
  \bibnamefont {Krasovskii}}, \bibinfo {author} {\bibfnamefont {E.~V.}\
  \bibnamefont {Chulkov}}, \bibinfo {author} {\bibfnamefont {H.}~\bibnamefont
  {Iwasawa}}, \bibinfo {author} {\bibfnamefont {T.}~\bibnamefont {Okuda}},
  \bibinfo {author} {\bibfnamefont {K.}~\bibnamefont {Shimada}}, \bibinfo
  {author} {\bibfnamefont {Y.}~\bibnamefont {Ueda}}, \bibinfo {author}
  {\bibfnamefont {H.}~\bibnamefont {Namatame}}, \ and\ \bibinfo {author}
  {\bibfnamefont {M.}~\bibnamefont {Taniguchi}},\ }\href {\doibase
  10.1103/PhysRevLett.105.076802} {\bibfield  {journal} {\bibinfo  {journal}
  {Phys. Rev. Lett.}\ }\textbf {\bibinfo {volume} {105}},\ \bibinfo {pages}
  {076802} (\bibinfo {year} {2010})}\BibitemShut {NoStop}%
\bibitem [{\citenamefont {Chang}\ \emph {et~al.}(2015)\citenamefont {Chang},
  \citenamefont {Tang}, \citenamefont {Feng}, \citenamefont {Li}, \citenamefont
  {Ma}, \citenamefont {Duan}, \citenamefont {He},\ and\ \citenamefont
  {Xue}}]{Chang2015}%
  \BibitemOpen
  \bibfield  {author} {\bibinfo {author} {\bibfnamefont {C.-Z.}\ \bibnamefont
  {Chang}}, \bibinfo {author} {\bibfnamefont {P.}~\bibnamefont {Tang}},
  \bibinfo {author} {\bibfnamefont {X.}~\bibnamefont {Feng}}, \bibinfo {author}
  {\bibfnamefont {K.}~\bibnamefont {Li}}, \bibinfo {author} {\bibfnamefont
  {X.-C.}\ \bibnamefont {Ma}}, \bibinfo {author} {\bibfnamefont
  {W.}~\bibnamefont {Duan}}, \bibinfo {author} {\bibfnamefont {K.}~\bibnamefont
  {He}}, \ and\ \bibinfo {author} {\bibfnamefont {Q.-K.}\ \bibnamefont {Xue}},\
  }\href {\doibase 10.1103/PhysRevLett.115.136801} {\bibfield  {journal}
  {\bibinfo  {journal} {Phys. Rev. Lett.}\ }\textbf {\bibinfo {volume} {115}},\
  \bibinfo {pages} {136801} (\bibinfo {year} {2015})}\BibitemShut {NoStop}%
\bibitem [{\citenamefont {E.Thareja}\ and\ \citenamefont
  {Vekhter}()}]{EThareja1}%
  \BibitemOpen
  \bibfield  {author} {\bibinfo {author} {\bibnamefont {E.Thareja}}\ and\
  \bibinfo {author} {\bibfnamefont {I.}~\bibnamefont {Vekhter}},\ }\href@noop
  {} {}\bibinfo {note} {Unpublished}\BibitemShut {NoStop}%
\bibitem [{\citenamefont {Dufouleur}\ \emph {et~al.}(2017)\citenamefont
  {Dufouleur}, \citenamefont {Veyrat}, \citenamefont {Dassonneville},
  \citenamefont {Nowka}, \citenamefont {Hampel}, \citenamefont {Leksin},
  \citenamefont {Eichler}, \citenamefont {Schmidt}, \citenamefont {Büchner},\
  and\ \citenamefont {Giraud}}]{Dufoul2017}%
  \BibitemOpen
  \bibfield  {author} {\bibinfo {author} {\bibfnamefont {J.}~\bibnamefont
  {Dufouleur}}, \bibinfo {author} {\bibfnamefont {L.}~\bibnamefont {Veyrat}},
  \bibinfo {author} {\bibfnamefont {B.}~\bibnamefont {Dassonneville}}, \bibinfo
  {author} {\bibfnamefont {C.}~\bibnamefont {Nowka}}, \bibinfo {author}
  {\bibfnamefont {S.}~\bibnamefont {Hampel}}, \bibinfo {author} {\bibfnamefont
  {P.}~\bibnamefont {Leksin}}, \bibinfo {author} {\bibfnamefont
  {B.}~\bibnamefont {Eichler}}, \bibinfo {author} {\bibfnamefont {O.~G.}\
  \bibnamefont {Schmidt}}, \bibinfo {author} {\bibfnamefont {B.}~\bibnamefont
  {Büchner}}, \ and\ \bibinfo {author} {\bibfnamefont {R.}~\bibnamefont
  {Giraud}},\ }\href {\doibase 10.1021/acs.nanolett.6b05051} {\bibfield
  {journal} {\bibinfo  {journal} {Nano Letters}\ }\textbf {\bibinfo {volume}
  {17}},\ \bibinfo {pages} {597} (\bibinfo {year} {2017})}\ \bibinfo {note}
  {}\BibitemShut {NoStop}%
\bibitem [{\citenamefont {Kamboj}\ \emph {et~al.}(2017)\citenamefont {Kamboj},
  \citenamefont {Singh}, \citenamefont {Ferrus}, \citenamefont {Beere},
  \citenamefont {Duffy}, \citenamefont {Hesjedal}, \citenamefont {Barnes},\
  and\ \citenamefont {Ritchie}}]{Kamboj2017}%
  \BibitemOpen
  \bibfield  {author} {\bibinfo {author} {\bibfnamefont {V.~S.}\ \bibnamefont
  {Kamboj}}, \bibinfo {author} {\bibfnamefont {A.}~\bibnamefont {Singh}},
  \bibinfo {author} {\bibfnamefont {T.}~\bibnamefont {Ferrus}}, \bibinfo
  {author} {\bibfnamefont {H.~E.}\ \bibnamefont {Beere}}, \bibinfo {author}
  {\bibfnamefont {L.~B.}\ \bibnamefont {Duffy}}, \bibinfo {author}
  {\bibfnamefont {T.}~\bibnamefont {Hesjedal}}, \bibinfo {author}
  {\bibfnamefont {C.~H.~W.}\ \bibnamefont {Barnes}}, \ and\ \bibinfo {author}
  {\bibfnamefont {D.~A.}\ \bibnamefont {Ritchie}},\ }\href {\doibase
  10.1021/acsphotonics.7b00492} {\bibfield  {journal} {\bibinfo  {journal} {ACS
  Photonics}\ }\textbf {\bibinfo {volume} {4}},\ \bibinfo {pages} {2711}
  (\bibinfo {year} {2017})}\ \Eprint
  {}
  {} \BibitemShut {NoStop}%
\bibitem [{\citenamefont {Chen}\ \emph {et~al.}(2013)\citenamefont {Chen},
  \citenamefont {Xie}, \citenamefont {Feng}, \citenamefont {Yi}, \citenamefont
  {Liang}, \citenamefont {He}, \citenamefont {Mou}, \citenamefont {He},
  \citenamefont {Peng}, \citenamefont {Liu}, \citenamefont {Liu}, \citenamefont
  {Zhao}, \citenamefont {Liu}, \citenamefont {Dong}, \citenamefont {Zhang},
  \citenamefont {Yu}, \citenamefont {Wang}, \citenamefont {Peng}, \citenamefont
  {Wang}, \citenamefont {Zhang}, \citenamefont {Yang}, \citenamefont {Chen},
  \citenamefont {Xu},\ and\ \citenamefont {Zhou}}]{Chen2013}%
  \BibitemOpen
  \bibfield  {author} {\bibinfo {author} {\bibfnamefont {C.}~\bibnamefont
  {Chen}}, \bibinfo {author} {\bibfnamefont {Z.}~\bibnamefont {Xie}}, \bibinfo
  {author} {\bibfnamefont {Y.}~\bibnamefont {Feng}}, \bibinfo {author}
  {\bibfnamefont {H.}~\bibnamefont {Yi}}, \bibinfo {author} {\bibfnamefont
  {A.}~\bibnamefont {Liang}}, \bibinfo {author} {\bibfnamefont
  {S.}~\bibnamefont {He}}, \bibinfo {author} {\bibfnamefont {D.}~\bibnamefont
  {Mou}}, \bibinfo {author} {\bibfnamefont {J.}~\bibnamefont {He}}, \bibinfo
  {author} {\bibfnamefont {Y.}~\bibnamefont {Peng}}, \bibinfo {author}
  {\bibfnamefont {X.}~\bibnamefont {Liu}}, \bibinfo {author} {\bibfnamefont
  {Y.}~\bibnamefont {Liu}}, \bibinfo {author} {\bibfnamefont {L.}~\bibnamefont
  {Zhao}}, \bibinfo {author} {\bibfnamefont {G.}~\bibnamefont {Liu}}, \bibinfo
  {author} {\bibfnamefont {X.}~\bibnamefont {Dong}}, \bibinfo {author}
  {\bibfnamefont {J.}~\bibnamefont {Zhang}}, \bibinfo {author} {\bibfnamefont
  {L.}~\bibnamefont {Yu}}, \bibinfo {author} {\bibfnamefont {X.}~\bibnamefont
  {Wang}}, \bibinfo {author} {\bibfnamefont {Q.}~\bibnamefont {Peng}}, \bibinfo
  {author} {\bibfnamefont {Z.}~\bibnamefont {Wang}}, \bibinfo {author}
  {\bibfnamefont {S.}~\bibnamefont {Zhang}}, \bibinfo {author} {\bibfnamefont
  {F.}~\bibnamefont {Yang}}, \bibinfo {author} {\bibfnamefont {C.}~\bibnamefont
  {Chen}}, \bibinfo {author} {\bibfnamefont {Z.}~\bibnamefont {Xu}}, \ and\
  \bibinfo {author} {\bibfnamefont {X.~J.}\ \bibnamefont {Zhou}},\ }\href
  {https://doi.org/10.1038/srep02411} {\bibfield  {journal} {\bibinfo
  {journal} {Scientific Reports}\ }\textbf {\bibinfo {volume} {3}},\ \bibinfo
  {pages} {2411 EP } (\bibinfo {year} {2013})}\ \bibinfo {note}
  {}\BibitemShut {NoStop}%
\bibitem [{\citenamefont {Kellner}\ \emph {et~al.}(2015)\citenamefont
  {Kellner}, \citenamefont {Eschbach}, \citenamefont {Kampmeier}, \citenamefont
  {Lanius}, \citenamefont {Młyńczak}, \citenamefont {Mussler}, \citenamefont
  {Holländer}, \citenamefont {Plucinski}, \citenamefont {Liebmann},
  \citenamefont {Grützmacher}, \citenamefont {Schneider},\ and\ \citenamefont
  {Morgenstern}}]{Kellner2015}%
  \BibitemOpen
  \bibfield  {author} {\bibinfo {author} {\bibfnamefont {J.}~\bibnamefont
  {Kellner}}, \bibinfo {author} {\bibfnamefont {M.}~\bibnamefont {Eschbach}},
  \bibinfo {author} {\bibfnamefont {J.}~\bibnamefont {Kampmeier}}, \bibinfo
  {author} {\bibfnamefont {M.}~\bibnamefont {Lanius}}, \bibinfo {author}
  {\bibfnamefont {E.}~\bibnamefont {Młyńczak}}, \bibinfo {author}
  {\bibfnamefont {G.}~\bibnamefont {Mussler}}, \bibinfo {author} {\bibfnamefont
  {B.}~\bibnamefont {Holländer}}, \bibinfo {author} {\bibfnamefont
  {L.}~\bibnamefont {Plucinski}}, \bibinfo {author} {\bibfnamefont
  {M.}~\bibnamefont {Liebmann}}, \bibinfo {author} {\bibfnamefont
  {D.}~\bibnamefont {Grützmacher}}, \bibinfo {author} {\bibfnamefont {C.~M.}\
  \bibnamefont {Schneider}}, \ and\ \bibinfo {author} {\bibfnamefont
  {M.}~\bibnamefont {Morgenstern}},\ }\href {\doibase 10.1063/1.4938394}
  {\bibfield  {journal} {\bibinfo  {journal} {Applied Physics Letters}\
  }\textbf {\bibinfo {volume} {107}},\ \bibinfo {pages} {251603} (\bibinfo
  {year} {2015})}\ \Eprint
  {}
  {} \BibitemShut {NoStop}%
\bibitem [{\citenamefont {Bianchi}\ \emph {et~al.}(2010)\citenamefont
  {Bianchi}, \citenamefont {Guan}, \citenamefont {Bao}, \citenamefont {Mi},
  \citenamefont {Iversen}, \citenamefont {King},\ and\ \citenamefont
  {Hofmann}}]{Bianchi2010}%
  \BibitemOpen
  \bibfield  {author} {\bibinfo {author} {\bibfnamefont {M.}~\bibnamefont
  {Bianchi}}, \bibinfo {author} {\bibfnamefont {D.}~\bibnamefont {Guan}},
  \bibinfo {author} {\bibfnamefont {S.}~\bibnamefont {Bao}}, \bibinfo {author}
  {\bibfnamefont {J.}~\bibnamefont {Mi}}, \bibinfo {author} {\bibfnamefont
  {B.~B.}\ \bibnamefont {Iversen}}, \bibinfo {author} {\bibfnamefont
  {P.~D.~C.}\ \bibnamefont {King}}, \ and\ \bibinfo {author} {\bibfnamefont
  {P.}~\bibnamefont {Hofmann}},\ }\href {\doibase 10.1038/ncomms1131}
  {\bibfield  {journal} {\bibinfo  {journal} {Nature Communications}\ }\textbf
  {\bibinfo {volume} {1}},\ \bibinfo {pages} {128} (\bibinfo {year}
  {2010})}\BibitemShut {NoStop}%
\bibitem [{\citenamefont {Fu}(2009)}]{Fu_hex2009}%
  \BibitemOpen
  \bibfield  {author} {\bibinfo {author} {\bibfnamefont {L.}~\bibnamefont
  {Fu}},\ }\href {\doibase 10.1103/PhysRevLett.103.266801} {\bibfield
  {journal} {\bibinfo  {journal} {Phys. Rev. Lett.}\ }\textbf {\bibinfo
  {volume} {103}},\ \bibinfo {pages} {266801} (\bibinfo {year}
  {2009})}\BibitemShut {NoStop}%
\bibitem [{\citenamefont {Nomura}\ \emph {et~al.}(2014)\citenamefont {Nomura},
  \citenamefont {Souma}, \citenamefont {Takayama}, \citenamefont {Sato},
  \citenamefont {Takahashi}, \citenamefont {Eto}, \citenamefont {Segawa},\ and\
  \citenamefont {Ando}}]{Nomura2014}%
  \BibitemOpen
  \bibfield  {author} {\bibinfo {author} {\bibfnamefont {M.}~\bibnamefont
  {Nomura}}, \bibinfo {author} {\bibfnamefont {S.}~\bibnamefont {Souma}},
  \bibinfo {author} {\bibfnamefont {A.}~\bibnamefont {Takayama}}, \bibinfo
  {author} {\bibfnamefont {T.}~\bibnamefont {Sato}}, \bibinfo {author}
  {\bibfnamefont {T.}~\bibnamefont {Takahashi}}, \bibinfo {author}
  {\bibfnamefont {K.}~\bibnamefont {Eto}}, \bibinfo {author} {\bibfnamefont
  {K.}~\bibnamefont {Segawa}}, \ and\ \bibinfo {author} {\bibfnamefont
  {Y.}~\bibnamefont {Ando}},\ }\href {\doibase 10.1103/PhysRevB.89.045134}
  {\bibfield  {journal} {\bibinfo  {journal} {Phys. Rev. B}\ }\textbf {\bibinfo
  {volume} {89}},\ \bibinfo {pages} {045134} (\bibinfo {year}
  {2014})}\BibitemShut {NoStop}%
\end{thebibliography}

%

\end{document}